\documentclass[aps,twocolumn,secnumarabic,nobalancelastpage,amsmath,amssymb,
nofootinbib]{revtex4}

\usepackage{graphics}      
\usepackage{graphicx}      
\usepackage{longtable}     
\usepackage{url}           
\usepackage{bm}            
\usepackage{color}

\begin{document}
\title{\textbf{Finite-time synchronization of tunnel
diode based chaotic oscillators}}
\author         {\bf Patrick Louodop$^{1}$, Hilaire Fotsin$^{1}$, Michaux
Kountchou$^{1}$,\\
\bf Elie B. Megam Ngouonkadi$^{1}$, Hilda A. Cerdeira$^{2}$ and Samuel
Bowong$^{3}$}

\affiliation    {$^1$ Laboratory of Electronics and Signal Processing
Faculty of Science, Department of Physics,
University of Dschang, P.O. Box 67 Dschang, Cameroon}
\affiliation    {$^2$ Instituto de F\'{i}sica Te\'{o}rica - UNESP, Universidade
Estadual Paulista,
Rua Dr. Bento Teobaldo Ferraz 271, Bloco II, Barra Funda,
 01140-070 S\~ao Paulo, Brazil}
\affiliation    {$^3$ Laboratory of Applied Mathematics,
Department of Mathematics and Computer Science, Faculty of Science,
University of Douala, P.O. Box 24157  Douala, Cameroon}

\date{\today}

\begin{abstract}
This paper addresses the problem of finite-time synchronization of tunnel diode
based chaotic oscillators. After a brief investigation of its chaotic dynamics,
we propose an active adaptive feedback coupling which
accomplishes the
synchronization of tunnel diode based chaotic systems with and
without the
presence of delay(s), basing ourselves on Lyapunov and on
Krasovskii-Lyapunov stability theories. This feedback coupling could be applied
to
many other chaotic systems. A finite horizon can be arbitrarily established by
ensuring that chaos synchronization is achieved at a
pre-established time. An
advantage of the proposed feedback coupling is that it is simple and easy to
implement. Both mathematical investigations and numerical simulations followed
by Pspice experiment are presented to show the feasibility of the proposed
method.
\end{abstract}
\maketitle

\section{Introduction}

Since chaotic systems were discovered, considerable interest arose in developing
and analyzing various systems that
exhibit chaos due to their importance in many fields of
sciences~\cite{bla,gra,lou}. In biology, epidemiology or in climatology,
investigating mathematical models is a part of the strategies used to better
comprehend the phenomenon~\cite{bla,gra,bow1,phil,earn,benhu}.
However, there exist
difficulties to
understand some of them because of the lack of data or the time
taken to produce reliable data. For example, in epidemiology, some deceases have
to be observed for a long period before they could be modeled~\cite{bow1}. To
turn around
these problems, electrical circuits (analogue computers) are built to allow the
observation of the effect of different parameters. Nevertheless, these
electrical
circuit have to be synchronized with the time history of data as much as
possible to be reliable. Hence, synchronization becomes an
important property to be studied.\\

Another characteristic of synchronization is its applications
in chaos based
cryptography~\cite{cru,bow2,fek}. Even if this problem can be
found in a number of papers,
most of them deal with infinite settling time~\cite{cru,bow2,fek}. If we
consider, for
example, the application of synchronization in secure communications, the range
of time during which the chaotic oscillators are not synchronized corresponds to
the range of time during which the encoded
message can unfortunately not be recovered or sent. More than a difficulty, this
is a
catastrophe in digital telecommunications, since the first bits of standardized
bit strings always contain signalization
data, i.e., the ``identity card'' of the message. Hence, it clearly appears that
the synchronization time has to be known, minimized,
so that the chaotic oscillators synchronize as fast as possible. In this
context, it is fair to say that there is a need to study finite-time chaos
synchronization. This characteristic is also helpful because of its advantages
in its applications in engineering. Hence, it is important to
investigate the finite-time stability of
nonlinear systems~\cite{yan,gui,laz,zup}.\\

Furthermore, even if delay complicates synchronization,
dynamical systems
with
delay(s) are abundant in nature. They occur in a wide variety of physical,
chemical, engineering, economics and biological systems and their networks.
There are many examples where delay plays an important role. Some of these
examples are listed and presented by M. Lakshamanan and D. V. Senthilkumar in
ref.~\cite{lak}. The mathematical description of delay dynamical systems will
naturally involve the delay parameter in some specified way. This can be in the
form of differential equations with delay or difference equations with delay and
so on~\cite{lak}. Delay differential equations (DDE) are given in many ways.
Again, we
refer the reader to ref.~\cite{lak} for more information.\\

In telecommunications, the delay notion is inescapable for
example because of the
relative distance between the transmitter and the receiver. Thus, the use of
fast signals is indicated. Presently a
solution is provided by electronic
components such as tunnel diodes. Developed in 1956 by L\'eo Esaki,
the tunnel diode is a nonlinear device used in very high frequencies amplifiers
with low noise and in micro-waves conception. The tunnel diode
based sources
can provide RF signals above 500 GHz.
Fundamental oscillations at 712 GHz from a resonant tunnel diode oscillator were
demonstrated by
Brown and co-workers in 1991~\cite{bro}. Recently, fundamental frequency
oscillations of a resonant tunnel
diode oscillator close to (831 GHz, 915 GHz) and above a terahertz (1.04 THz) at
room temperature
were reported by Suzuki et al.~\cite{suz} -- more details can
be found in
the PhD thesis of Wang,
Liquan (2011) of University of Glasgow~\cite{liq} and
references therein. Considering the
frequency capability of the tunnel diode, it can be used as an
efficient source for chaotic signals in
 wireless communications~\cite{wir}. In particular
Hai-Peng Ren et al. show that, for some particular chaotic signals the Lyapunov
exponents remain unaltered through multipath propagation, thus making these
systems able to sustain
communications with chaotic signals.
  However, experimental observation of generalized
synchronization phenomena
in microwave oscillators have been done by B. S. Dmitriev et
al.~\cite{fir} and their application to secure communications has been carried
out
by A. A. Koronovskii et al.~\cite{sec}. Nevertheless, the synchronization
schemes used in the mentioned references~\cite{wir,fir,sec} are dealing with
asymptotical synchronization and can only guarantee that two systems achieve
synchronization as time tends to infinity, while in real-world
applications, one always expects that two systems achieve synchronization within
a finite and/or predetermined time. Besides, finite-time
synchronization
has been confirmed to have better robustness against parametric and external
disturbances than the asymptotical one~\cite{yon,hua}. Hereby
the importance to study finite-time
synchronization of chaotic systems. The use of tunnel diodes to
construct chaotic systems was also reported by A. Pikovskii~\cite{pik},
A. Fradkov~\cite{mar} and others. For the purpose of this work, we base
ourselves on the circuit proposed by A. Fradkov~\cite{mar} where we
replace the
used
tunnel diode by the one with reference number 1N3858 for which
the Pspice model is
available in Internet.\\

In this paper we study an active adaptive finite-time synchronization of tunnel
diodes based chaotic oscillators with (first case) and without
(second and third cases)
time-delay in both the drive and response systems.
Therefore, we will investigate the adaptive synchronization of chaotic systems
using just one master state variable as output to construct the nonlinear
feedback coupling. A robust adaptive response system will be
therefore designed to always globally synchronize the given driven tunnel diode
based oscillator. These results are of significant interest to infer
relationships between nonlinear feedback coupling, time-delay and finite-time
synchronization. To the best of our knowledge, this active
adaptive
finite-time synchronization of chaotic systems using a nonlinear feedback
coupling has not yet being reported in the literature.\\

The manuscript is organized as follows: In section 2, the circuit model and its
chaotic dynamics is investigated. Section 3 deals with the adaptive
synchronization problem. In this section the active controller is designed and
the theoretical settling time in the case of systems without delay is developed.
 Numerical simulations are performed to show the effectiveness
of the
synchronization scheme. In section 4, we investigate the cases of delayed
systems considering two cases. The theoretical settling is also obtained and the
numerical results are presented graphically to prove the
effectiveness
of the scheme. In the last section we present the conclusions.


\section{The model and its properties}
\noindent

\begin{figure}[htp]
 \begin{minipage}[b]{8cm}
      \begin{center}
    \includegraphics[scale=0.43]{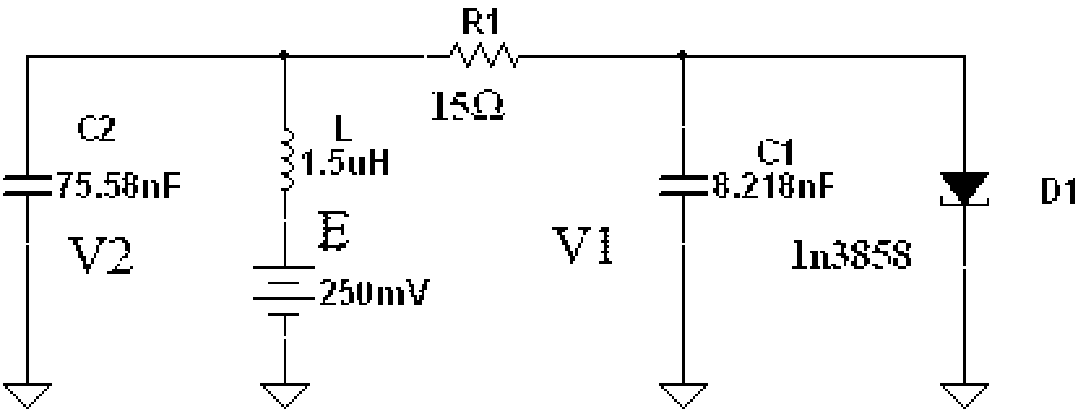}(a)
 \includegraphics[scale=0.21]{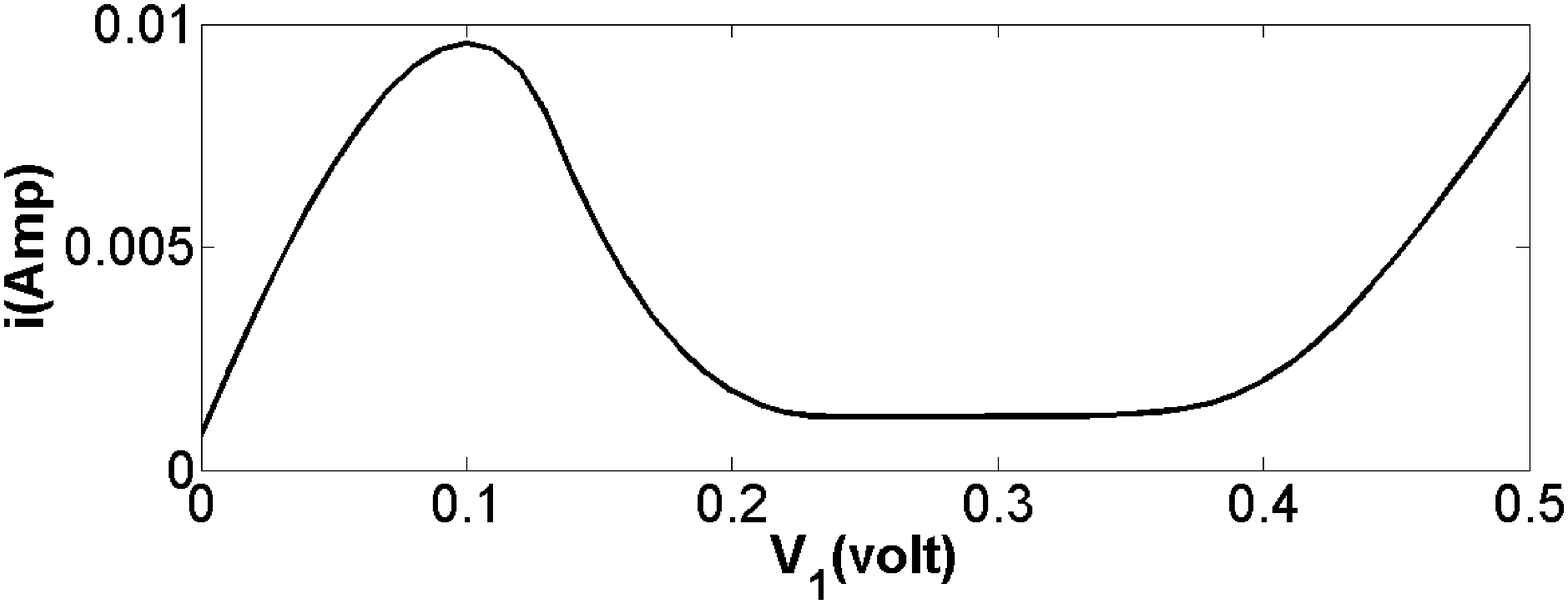}(b)
    \caption{(a) Circuit diagram and (b) current voltage characteristic of
tunnel diode 1N3858. }
     \label{cir}
     \end{center}
     \end{minipage}
     \end{figure}

Before starting to analyze adaptive synchronization and delay,
we introduce the elements of our circuit. Its electronic
diagram is
given on Fig.~(\ref{cir}.a) where the
tunnel diode model is the 1N3858 with the current-voltage characteristic given
on Fig.~(\ref{cir}.b). The chaotic circuit is a  R-L-C oscillator where $E$ is a
continuous voltage source.

The circuit diagram presents two capacitors $C_1$ and $C_2$ and an inductor $L$.
The nonlinearity is introduced by the tunnel diode which has its current-voltage
characteristic depicted in Fig.~(\ref{cir}.b).
As an approximation of this current-voltage characteristic, we
use the
following relation~\cite{mar}:
\begin{equation} \label{nonl}
 i(V_1)=a_1(V_1-b)^3-a_2(V_1-b)+a_3,
\end{equation}
where $a_1$, $a_2$ and $a_3$ are positive constants. The circuit
can be described by the following system
of differential equations:
\begin{equation}
 \label{model}
\left\{\begin{array}{lcl}
\dot{x}_1(\tau)&=&\alpha [x_2-x_1-rf(x_1)],\\
\\
\dot{x}_2(\tau)&=&\beta(x_1-x_2+r x_3),\\
\\
\dot{x}_3(\tau)&=&\gamma(E-x_2),
       \end{array}
\right.
\end{equation}
where
$$\begin{array}{ll}c=\displaystyle\frac{C_2}{C_1},\qquad r=R_1,\qquad
q=\displaystyle\frac{r}{Lw},\qquad \tau=wt,\\
\varrho=1\,V,\qquad I=1\,A,\qquad  x_1=\displaystyle\frac{V_1}{\varrho},\qquad
x_2=\displaystyle\frac{V_2}{\varrho},\\
x_3=\displaystyle\frac{i}{I},\qquad \alpha=\displaystyle\frac{c}{q},
\beta=\displaystyle\frac{1}{q},\qquad \gamma=\displaystyle\frac{q}{r}
\qquad\mbox{and}\\ f(x_1)=a_1(x_1-b)^3-a_2(x_1-b)+a_3,
\end{array}$$ where $V_1$ and $V_2$ are the voltages at landmarks of $C_1$ and
$C_2$ respectively, $i$ is the current which flows through
the inductor and $w$ is a constant. Note that Eq.~(\ref{model}) is similar to
the so called modified
Chua's circuit~\cite{yas}.

%

\begin{figure}[htp]
 \begin{minipage}[b]{8cm}
      \begin{center}
    \includegraphics[scale=0.25]{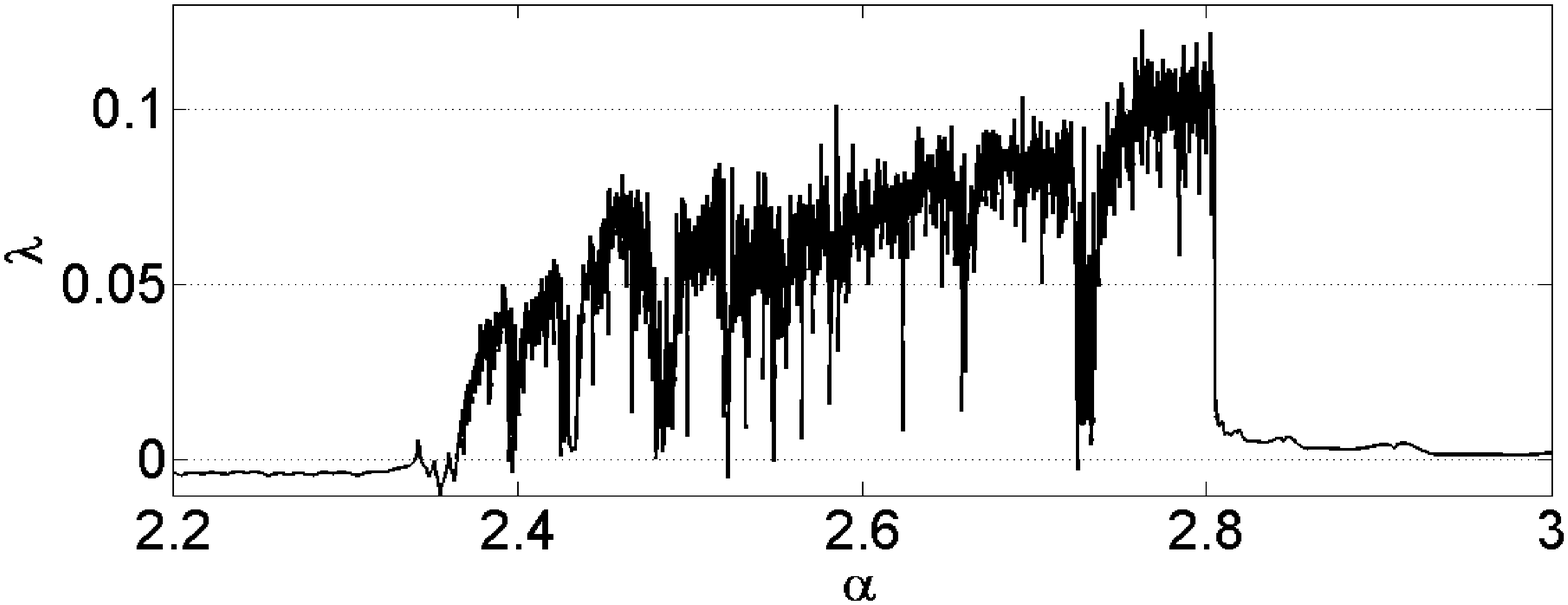}(a)
 \includegraphics[scale=0.25]{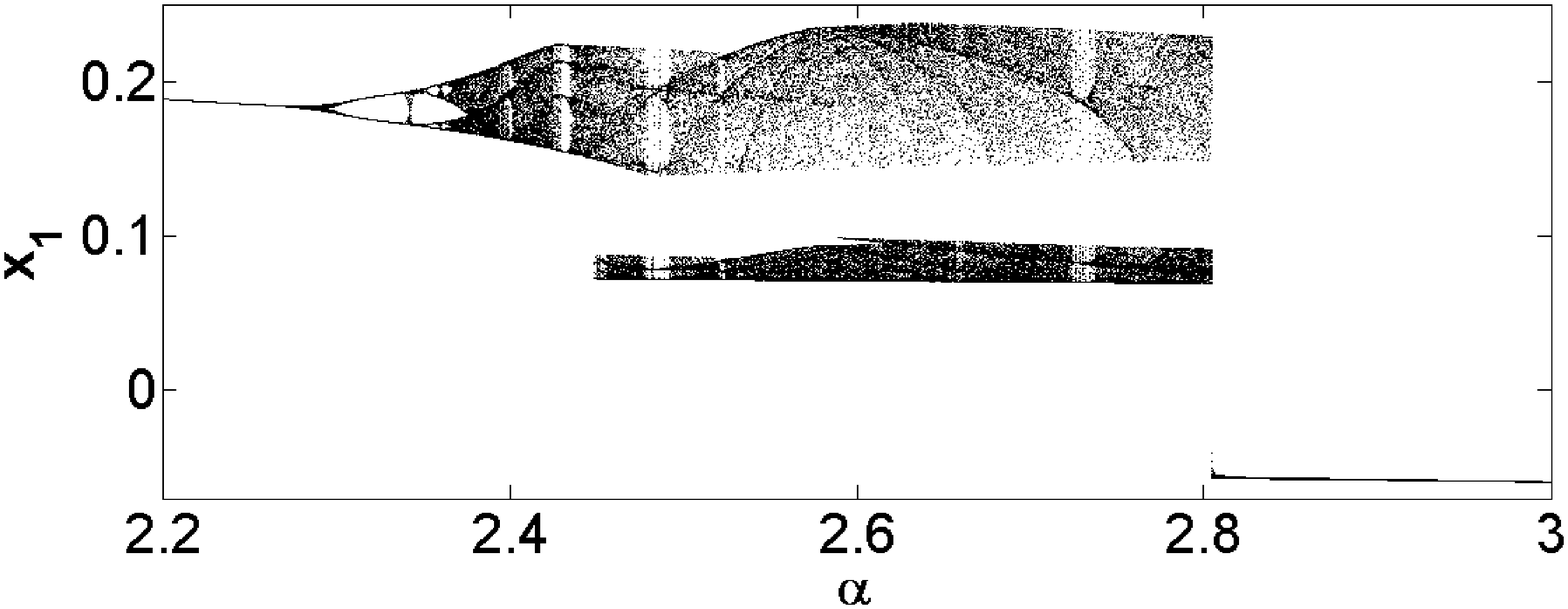}(b)
    \caption{(a) One dimensional Lyapunov exponent and (b) bifurcation diagram
when $E=0.253208$ and $q=3.6$. }
     \label{lya}
     \end{center}
     \end{minipage}
     \end{figure}

Before entering into the description of the system under study
we present the chaotic behavior of the
circuit due to the influence of parameter $\alpha$ on the evolution of the
one-dimensional Lyapunov exponent Fig.~(\ref{lya}.a) and on bifurcation diagram
Fig.~(\ref{lya}.b)
These two graphs allow us to determine the value of
the control parameter, leading to a chaotic behavior of the system
through period doubling.
Fig.~(\ref{lya}.a) shows that for some values of the parameter $\alpha$ the
system Eq.~(\ref{model}) has a positive Lyapunov exponents.

\begin{figure}[htp]
 \begin{minipage}[b]{8cm}
      \begin{center}
    \includegraphics[scale=0.15]{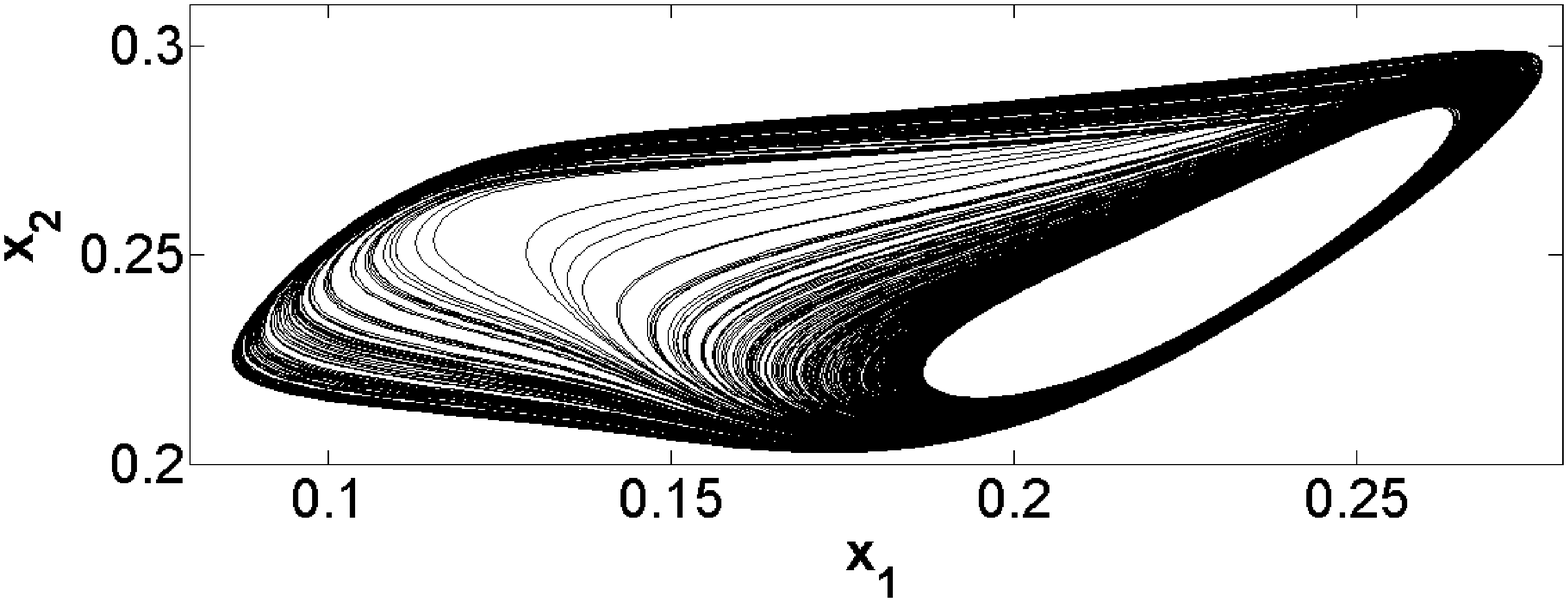}(a)
 \includegraphics[scale=0.15]{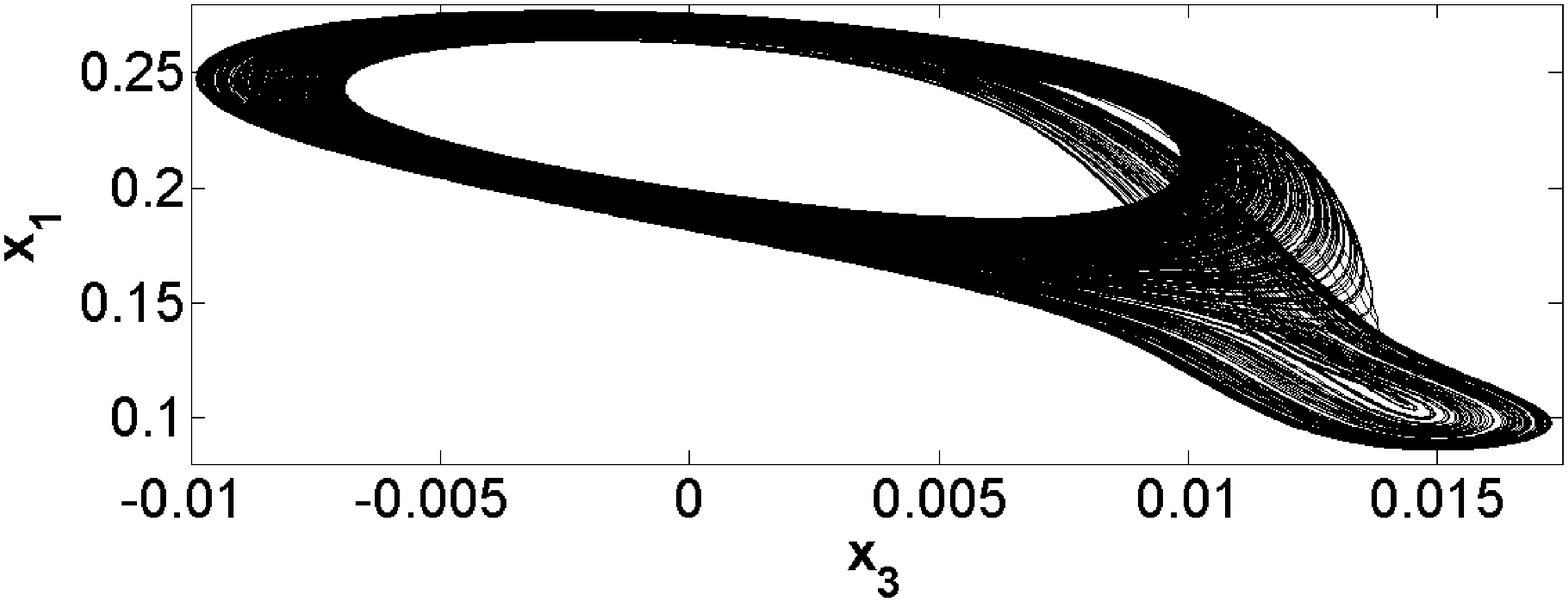}(b)
    \caption{ Chaotic attractor obtained through Pspice simulation of the
circuit of Fig.1(a).}
     \label{attp}
     \end{center}
     \end{minipage}
     \end{figure}

Figs.~(\ref{attp}.a and b) show the graphs $x_2$ vs $x_1$ and $x_1$ vs $x_3$
obtained from Pspice simulations of the circuit of Fig.~(\ref{cir}.a). An approximate behavior is displayed with the following
selection of parameters $c=8.4$, $q=3.35$, $E=0.25$, $r=16$, $a_1 =1.3242872$,
$a_2=0.06922314$, $a_3=0.005388$ and $b=0.167$ and initial conditions $x_1(0)=0.15$ , $x_2(0)=0.27$ and $x_3(0)=0.008$.

\begin{figure}[htp]
 \begin{minipage}[b]{8cm}
      \begin{center}
     \includegraphics[scale=0.15]{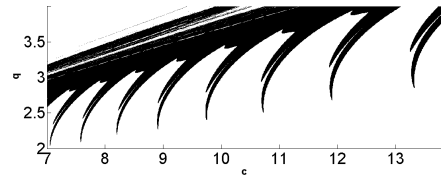}
    \caption{Pair $(c,q)$ for which chaos occurs with $E=0.25$.}
     \label{bas}
     \end{center}
     \end{minipage}
     \end{figure}
\begin{figure}[htp]
 \begin{minipage}[b]{8cm}
      \begin{center}
    \includegraphics[scale=0.110]{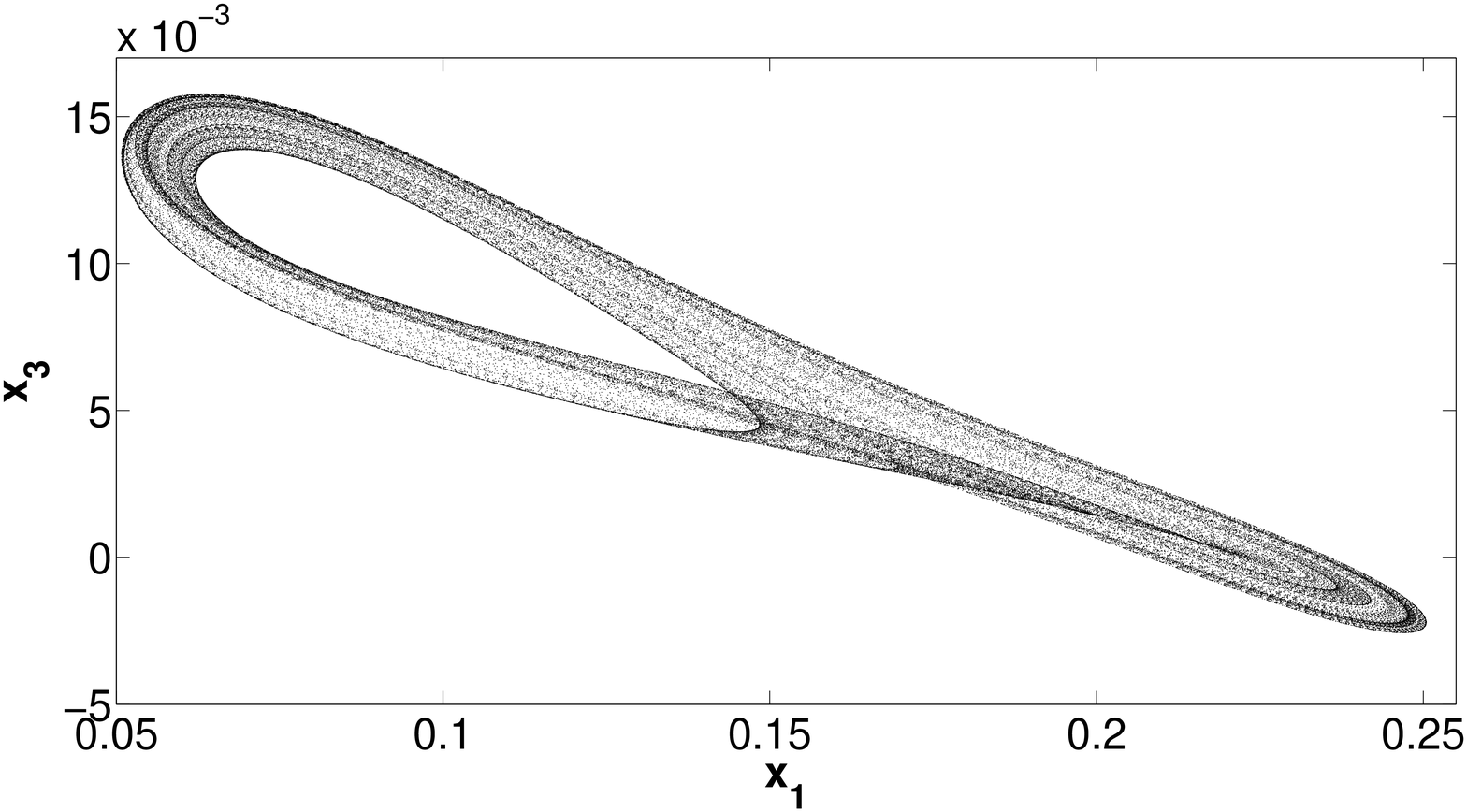}(a)
 \includegraphics[scale=0.110]{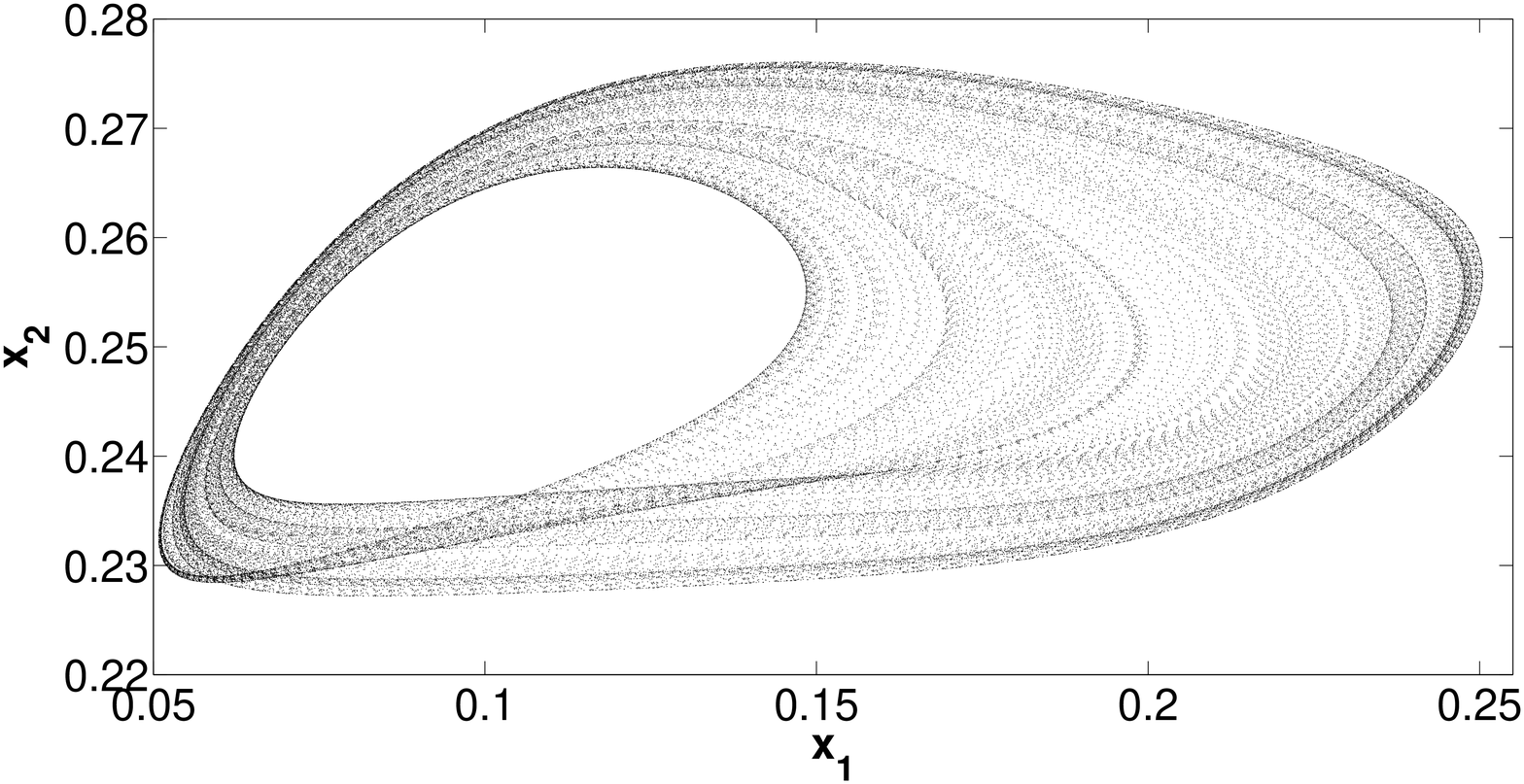}(b)
    \caption{ Chaotic attractor obtained through numerical simulation of
Eq.~(2).}
     \label{attm}
     \end{center}
     \end{minipage}
     \end{figure}

In Fig.~(\ref{bas}) the space $(c,q)$ for
which the system behaves chaotically is shown in black.
With $\alpha=2.507462687$ and the values of the parameters
given above, basing ourselves
on Fig.~\ref{bas}
, it follows that, the Lyapunov exponent is positive and
the system shows chaotic behavior (Fig.~\ref{attm}).

\begin{figure}[htp]
 \begin{minipage}[b]{8cm}
      \begin{center}
    \includegraphics[scale=0.12]{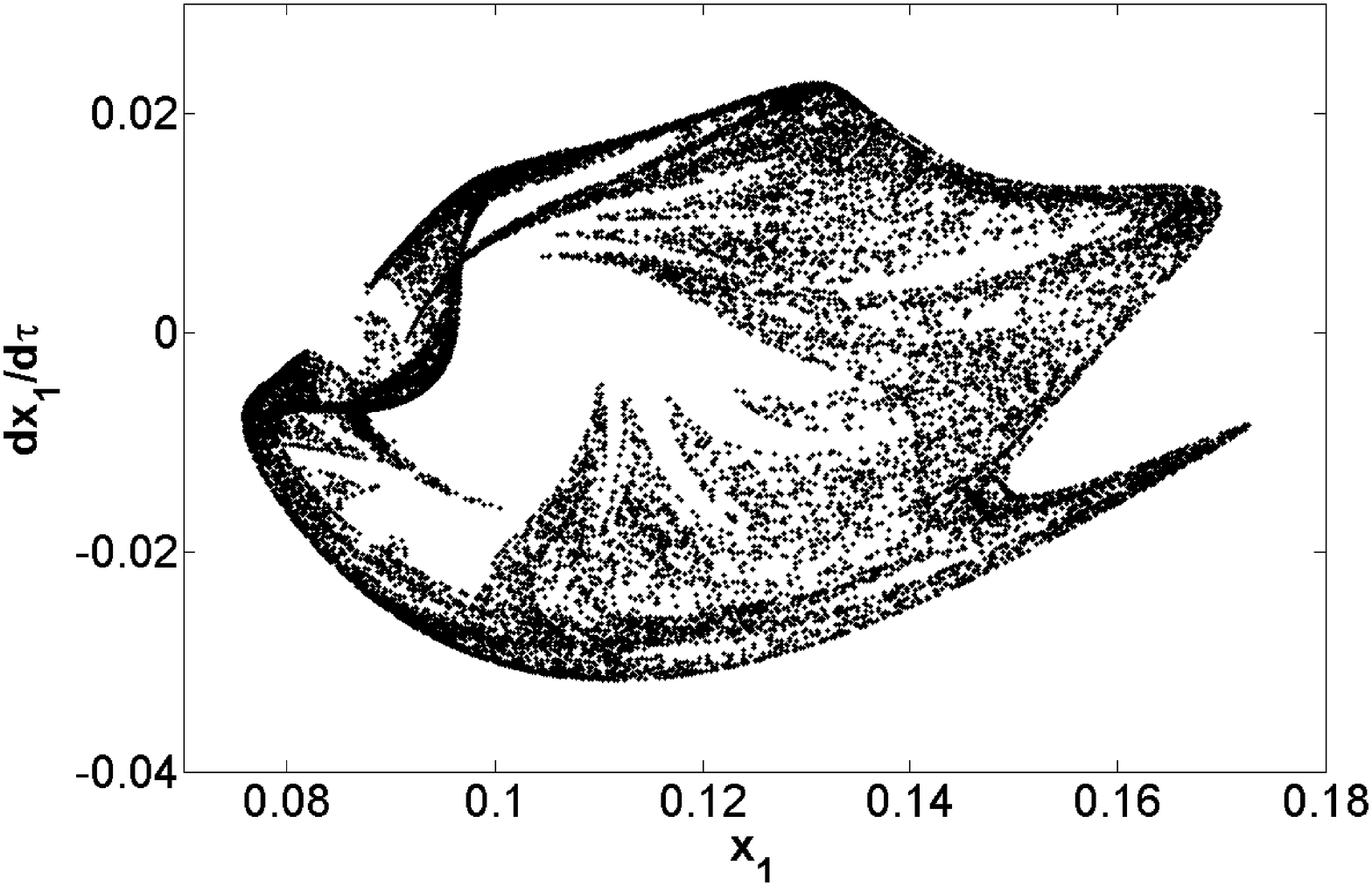}(a)
 \includegraphics[scale=0.12]{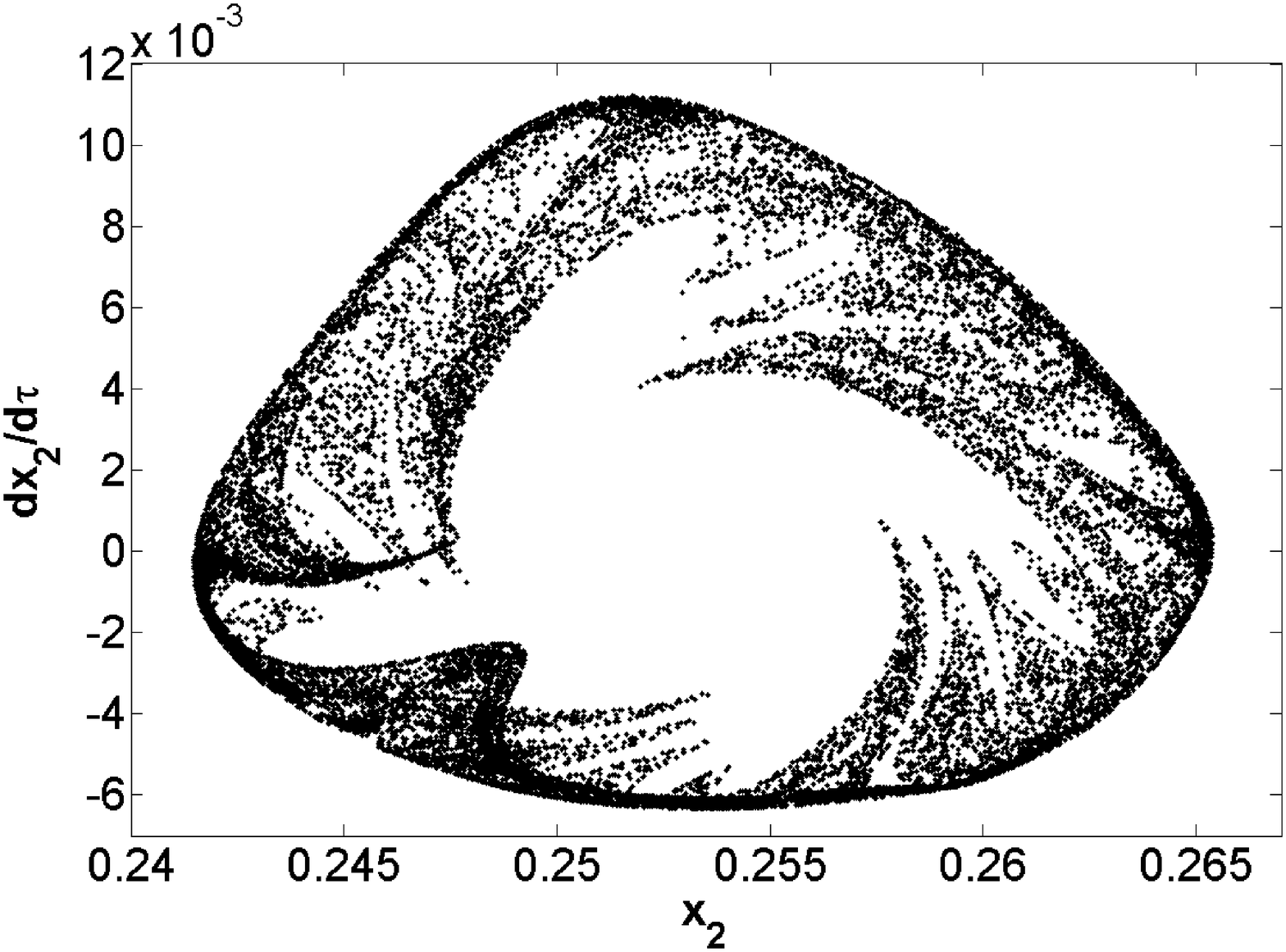}(b)
 \includegraphics[scale=0.12]{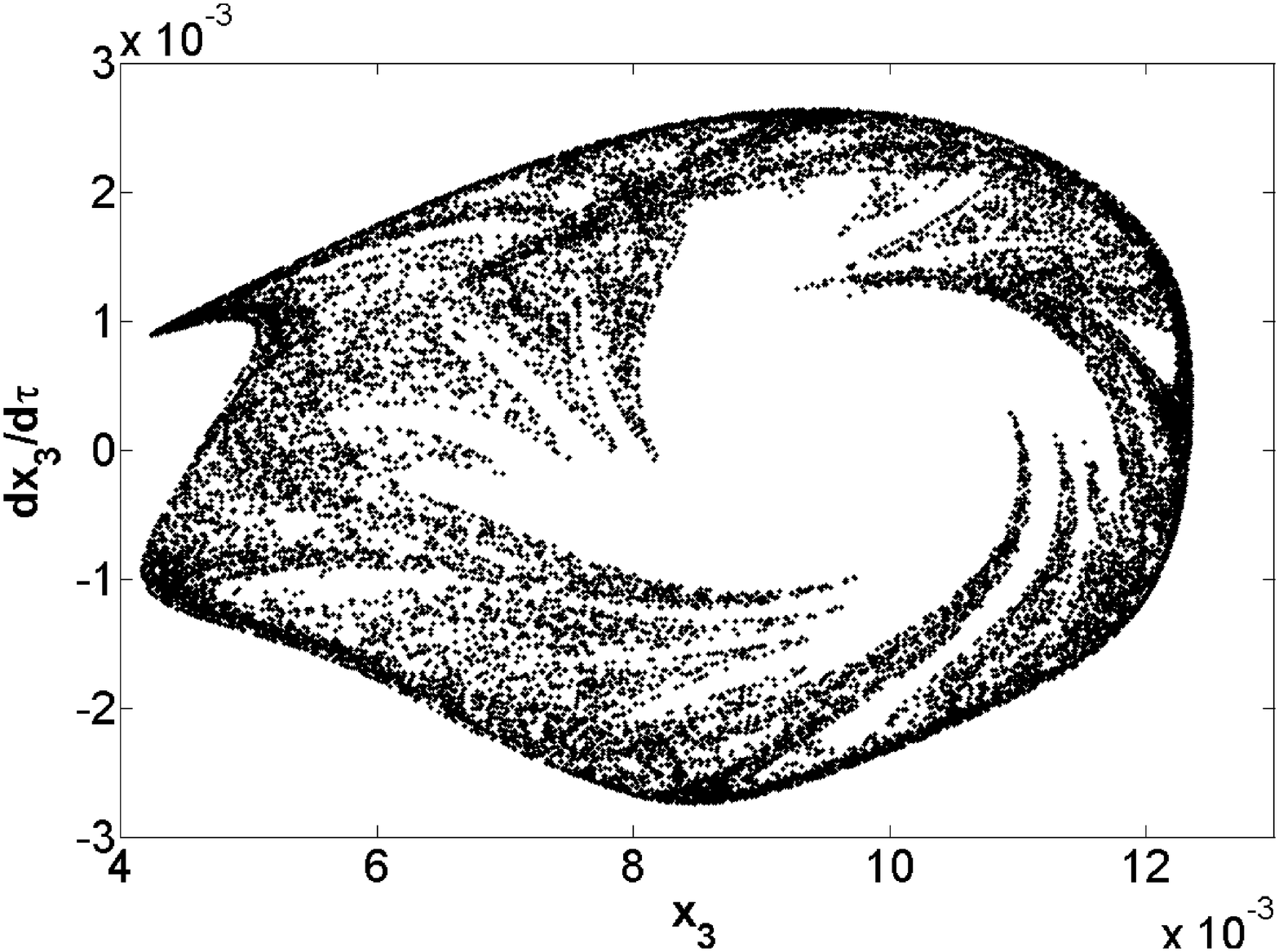}(c)
    \caption{Poincare sections with $E=0.253205$, $c=8.846$ and $q=3.6$.}
     \label{poin}
     \end{center}
     \end{minipage}
     \end{figure}

    The graphs of Figs.~(\ref{poin}) show the Poincare sections which help to
prove the chaotic behavior of the model (Eq.~(\ref{model})).

\section{Adaptive synchronization}
\noindent

 The starting point of the control law design is always a mathematical model of
the real system which describes the system response to various control inputs.
It is usually assumed that such a model is known rather completely. However, in
practice the model describes the real system with some degree of uncertainty. To
cope with this uncertainty in such cases, methods of the adaptive control theory
can be employed. The concept of chaos synchronization emerged much later, not
until the gradual realization of the usefulness of chaos by scientists and
engineers. Chaotic signals are usually broad band and noise-like. Because of
these properties, synchronized chaotic systems can be used as cipher generators
for secure communication.\\
In this section, we investigate two identical systems of the
type described in the previous section where the system
drive $x_i(\tau), i=1,2,3$ drives the response system $y_i(\tau), i=1,2,3$.

\subsection{Design of the nonlinear controller}
\noindent

Here we investigate the finite-time synchronization of tunnel
diode based chaotic oscillators described by Eq.~(\ref{model}).
%
For this purpose we consider the corresponding response system
described by
\begin{equation}
\label{cresponse} \left\{ \begin{array}{lcl}
\dot{y}_1&=&\alpha \left[y_2-y_1-r f(y_1)\right]-\zeta k~sign(y_1-x_1)-\zeta
u(\tau),\\
\\
\dot{y}_2&=&\beta\left[y_1-y_2+ r y_3-2(y_1-x_1)\right],\\
\\
\dot{y}_3&=&\gamma \left(E-y_2\right),\end{array} \right.
\end{equation}
where $k$ and $\zeta$ are positive constant gains to be defined by the
designer. The term $-2(y_1-x_1)$ added in the $y_2$-axis stabilizes the slave
system. The nonlinear controller is expressed as $-\zeta k~sign(y_1-x_1)-\zeta
u(\tau)$ where $u(\tau)$ is the adaptive feedback coupling designed to achieve
finite-time synchronization. All parts of the used controller contribute to
synchronize both drive-response systems at an established time
and also to
stabilize the response dynamics.\\

Let us define the synchronization error as follows

 \begin{equation} \label{err}
  \begin{array}{lcl}
e_i \left( \tau\right) &=& y_{i}\left( \tau\right)- x_i\left( \tau\right),\qquad
\mbox{with }\qquad i=1, 2, 3.
\end{array}
\end{equation}

Subtracting Eq.~(\ref{model}) from Eq.~(\ref{cresponse}) and using
Eq.~(\ref{err}), we obtain:

\begin{equation}\label{errd} \left\{ \begin{array}{lcl}
\dot{e}_1&=&\alpha \left[e_2-e_1-\phi_0(x_1,y_1)\right]-\zeta k sign(e_1)-\zeta
u(\tau),\\
\\
\dot{e}_2&=&\beta\left[-e_1-e_2+ r e_3\right],\\
\\
\dot{e}_3&=&-\gamma e_2,\end{array} \right.
\end{equation}
where $\phi_0(x_1,y_1)=r \left(f(y_1)-f(x_1)\right)$.\\
The synchronization problem can be stated as follows: considering the
transmitter Eq.~(\ref{model}) and the receiver Eq.~(\ref{cresponse}) with any
initial conditions, $e_1(0)=y_1(0)- x_1(0)$, $e_2(0)=y_2(0)- x_2(0)$,
$e_3(0)=y_3(0)- x_3(0)$, it is our aim to design the form of
the function $u(\tau)$ which
synchronizes the orbits of both the transmitter and the response systems and
thus provides the stabilization of Eq.~(\ref{errd}) at an established
finite-time $\tau_s$, i.e.,
\begin{equation}
\label{limit}\begin{array}{lcl}
\lim\limits_{t\rightarrow \tau_s}x_i(\tau)= y_i(\tau),\qquad i=1,2,3,\\
\\
\lim\limits_{\tau \rightarrow \tau_s}\Arrowvert e(\tau )\Arrowvert=0.
\end{array}
\end{equation}
where $\tau_s$ is the settling time. We consider that, the function
$\phi_0(x_1,y_1)$ respects the Lipschitz condition. I.e., it exists a positive
constant $\chi_0$ such that $|\phi_0(x_1,y_1)|\leq \chi_0 |e_1(\tau)|$.

\subsection{Main results}
\noindent

The global finite-time synchronization is achieved when the Lyapunov function
candidate is proper~\cite{zup}, namely, the time derivative of the used Lyapunov
function is bounded by a negative constant.\\
Let us consider the following candidate for the Lyapunov  function~\cite{lia}

\begin{equation}\label{lyap}
V
=\displaystyle\frac{1}{2}\left(\displaystyle\frac{e_1^2}{\alpha}
+\displaystyle\frac{e_2^2}{\beta}
+\displaystyle\frac{r e_3^2}{\gamma}\right)+| u(\tau) |.
\end{equation}
The time derivative along the trajectories of the system of Eqs.~(\ref{errd})
yields

\begin{equation}
\label{lyd1} \begin{array}{lcl}
\dot{V} &=&-e_1^2-e_2^2-\phi_0(x_1,y_1) e_1-\displaystyle\frac{\zeta k}{\alpha}
|e_1|-\displaystyle\frac{\zeta u(\tau) }{\alpha}e_1\\
&+&sign(u)\dot{u}(\tau),\\
\\
&\leq& -e_1^2-e_2^2+|\phi_0(x_1,y_1)||e_1| - \displaystyle\frac{\zeta k}{\alpha}
|e_1|-\displaystyle\frac{\zeta u(\tau) }{\alpha} e_1 \\
&+&sign(u)\dot{u}(\tau),\\
\\
&\leq& -\left(1-\chi_0\right)e_1^2-e_2^2- \displaystyle\frac{\zeta k}{\alpha}
|e_1|-\displaystyle\frac{\zeta u(\tau) }{\alpha} e_1 \\
&+&sign(u)\dot{u}(\tau).
\end{array}
\end{equation}
Letting $\chi_0 < 1$ and $\zeta = \alpha$,
\begin{equation}
\label{lyd2} \begin{array}{lcl}
\dot{V} &\leq& - k |e_1|- u(\tau) e_1 +sign(u)\dot{u}(\tau).
\end{array}
\end{equation}

Considering the finite-time stability theory applied in~\cite{zup}, it follows
that,
there exists a certain positive constant $p$ such that the following relation
yields

\begin{equation}\label{lyd3}
\dot{V} \leq - p.
\end{equation}

If the objective is reached, the theoretical finite-time for
synchronization is
obtained by integrating Eq.~(\ref{lyd3}) from $0$ to $\tau_s$. Thus, one has

\begin{equation}\label{ctos}
\tau_s \leq \displaystyle\frac{1}{2
p}\left(\displaystyle\frac{e_1^2(0)}{\alpha}+\displaystyle\frac{e_2^2(0)}{\beta}
+\displaystyle\frac{r e_3^2(0)}{\gamma}\right)+\displaystyle\frac{| u(0) |}{p}.
\end{equation}

Therefore, in accordance with Eq.~(\ref{lyd3}), the controller
$u(\tau)$ is designed
through the following relation

\begin{equation}\label{cont}
\dot{u}= sign(u)\left(k|e_1|+ u e_1- p \right),
\end{equation}

and we construct the slave system Eq.~(\ref{cresponse}) as
follows

\begin{equation}
\label{rescont} \left\{ \begin{array}{lcl}
\dot{y}_1&=&\alpha \left[y_2-y_1-r f(y_1)\right]-\zeta
k~sign(y_1-x_1)-\zeta u,\\
\\
\dot{y}_2&=&\beta\left[y_1-y_2-r y_3-2(y_1-x_1)\right],\\
\\
\dot{y}_3&=&\gamma \left[E-y_2\right],\\
\\
\dot{u}&=& sign(u)\left(k|e_1|+ u e_1- p \right).
\end{array} \right.
\end{equation}

\subsection{Numerical results}
\noindent

\begin{figure}[htp]
 \begin{minipage}[b]{8cm}
      \begin{center}
    \includegraphics[scale=0.18]{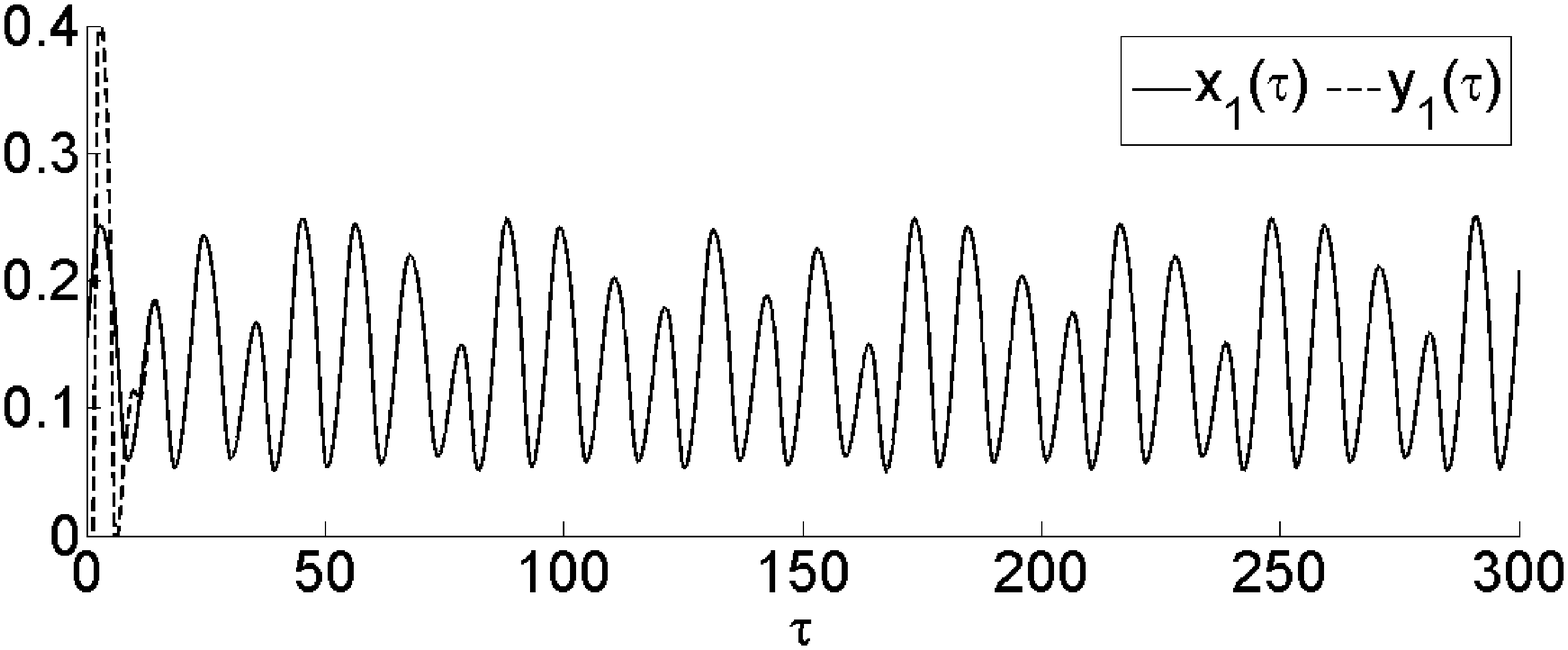}(a)
        \includegraphics[scale=0.18]{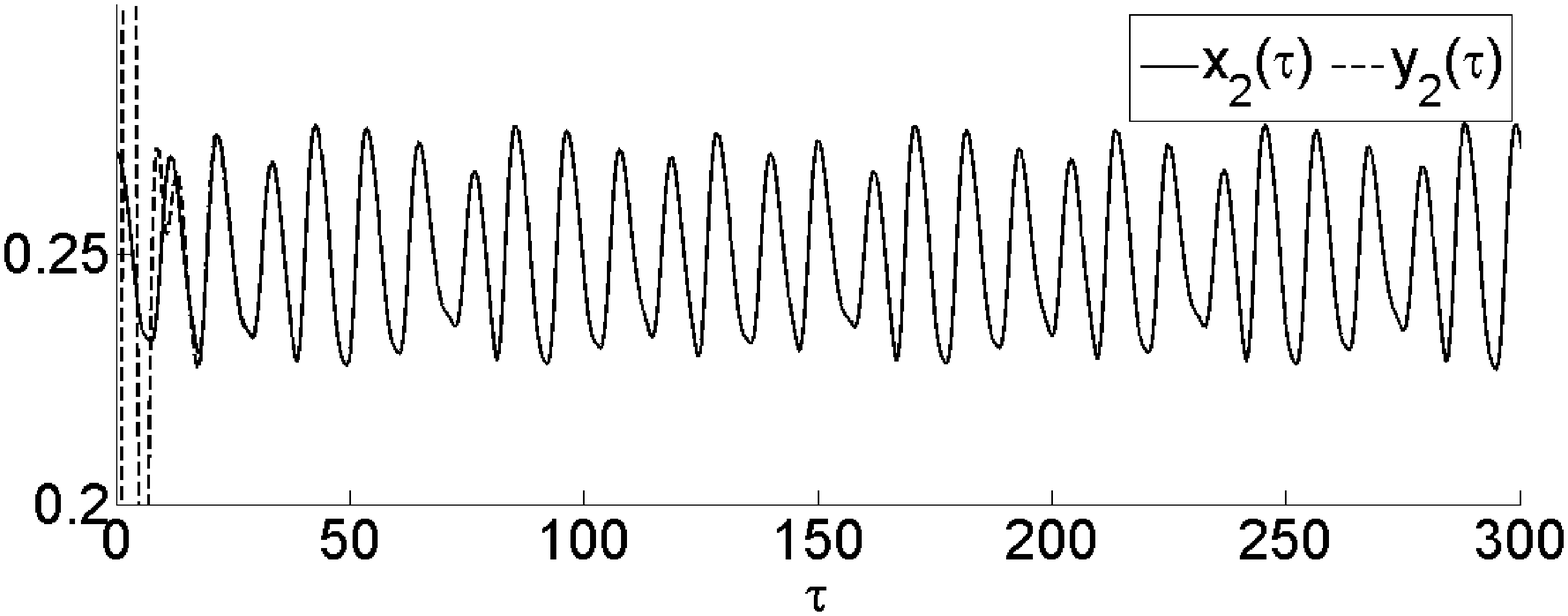}(b)
        \includegraphics[scale=0.18]{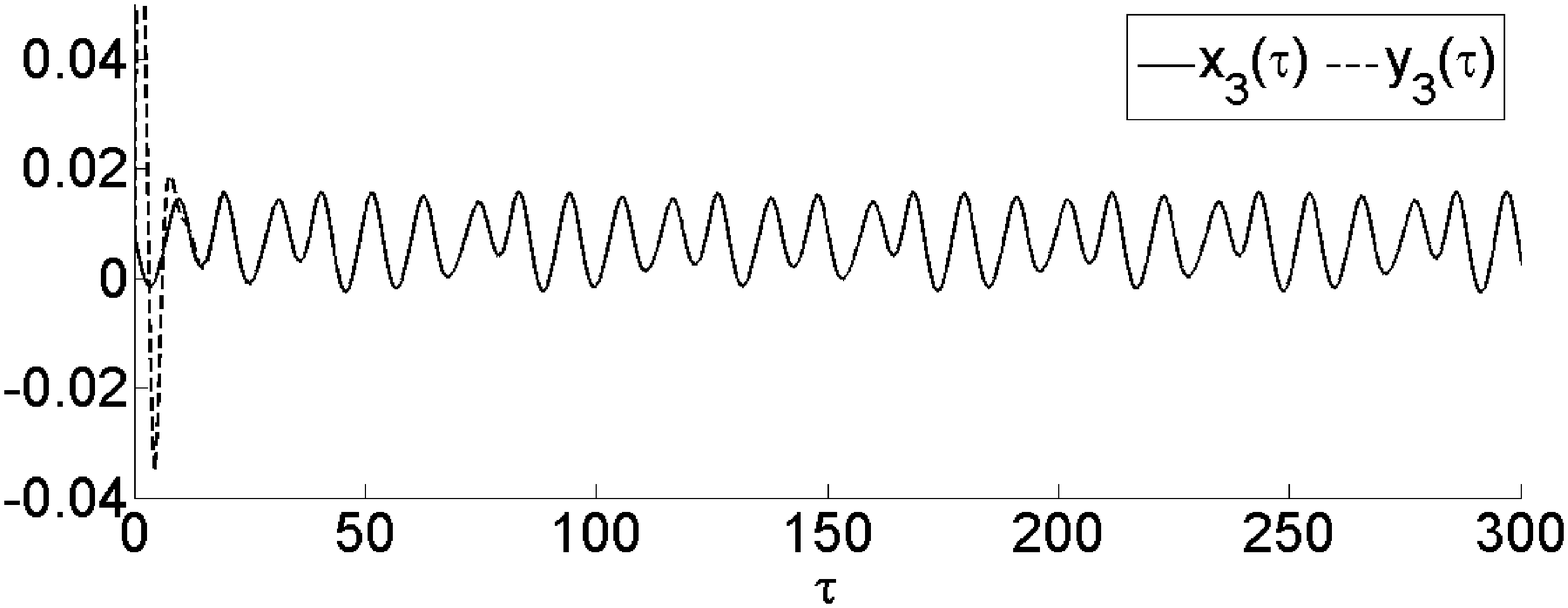}(c)
    \caption{Time histories of state variables, solid line for drive system and
dashed line for response system.}
     \label{vi}
     \end{center}
     \end{minipage}
     \end{figure}

\begin{figure}[htp]
 \begin{minipage}[b]{8cm}
      \begin{center}
    \includegraphics[scale=0.18]{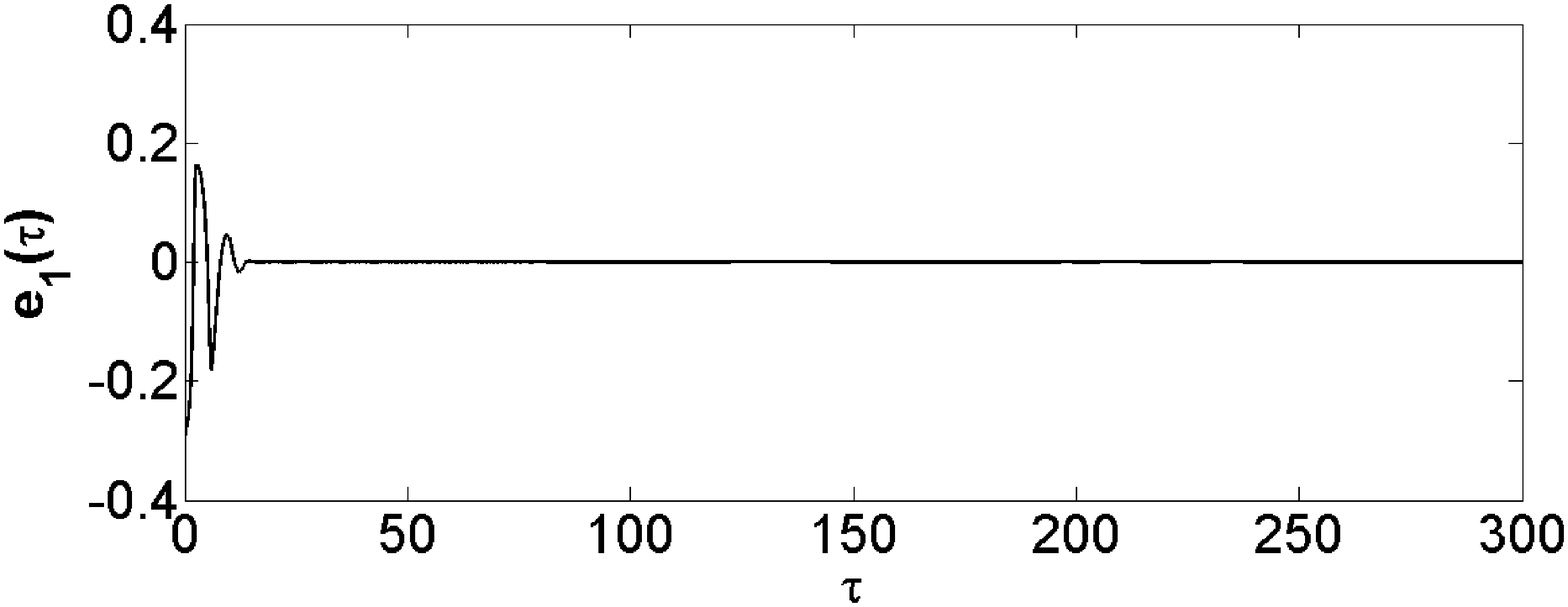}(a)
        \includegraphics[scale=0.18]{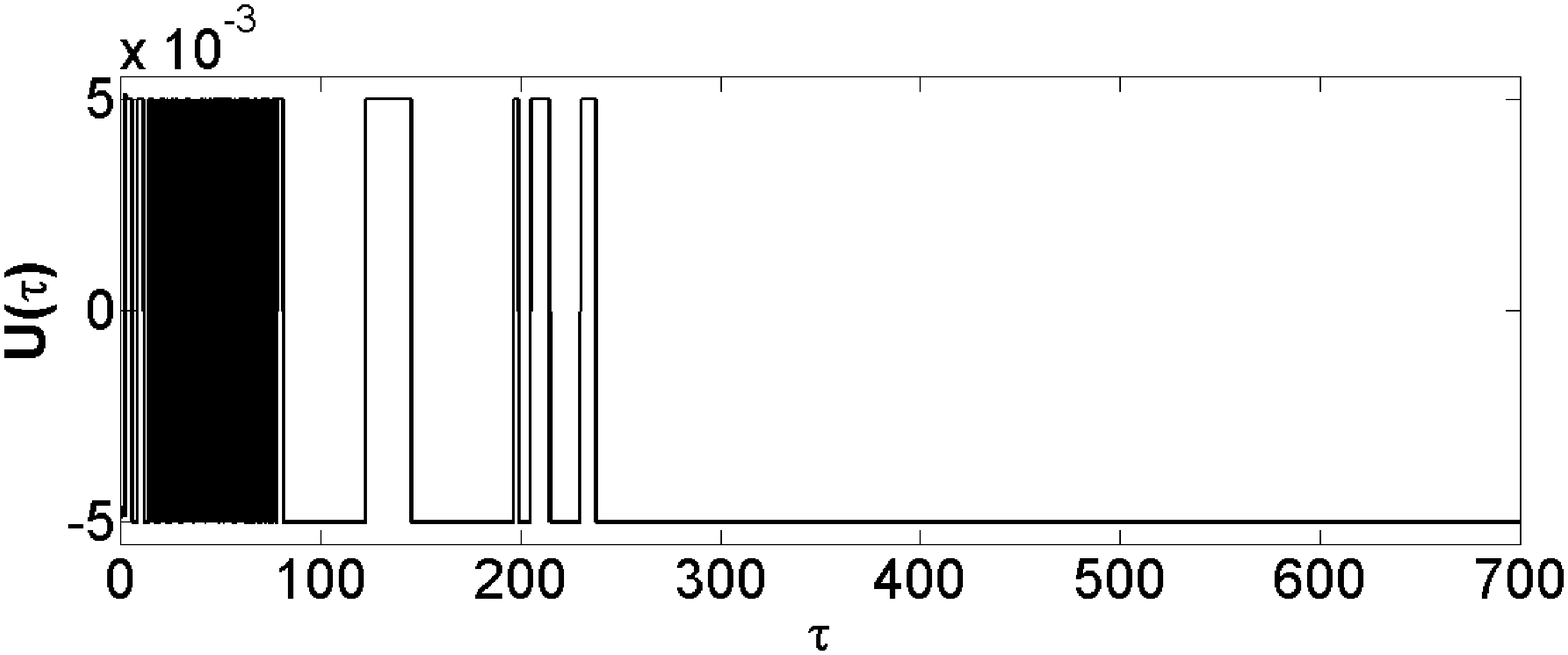}(b)
    \caption{Time histories of $e_1(\tau)$ (a) and of the controller $u(\tau)$
(b).}
     \label{eu}
     \end{center}
     \end{minipage}
     \end{figure}

To illustrate the effectiveness of
the proposed scheme we present some numerical results for the
initial conditions given by
$(x_{1}(0),x_{2}(0),x_{3}(0))=(0.15, 0.27,
0.008)$ and $(y_{1}(0),y_{2}(0),y_{3}(0))
=(-0.15, -0.27, -0.008)$, the parameters $p=0.001$, $k=0.005$
and the initial
condition of controller $u(0)=0.0001$.
The integration time step was taken as $10^{-4}$. For such values, the
theoretical settling time is determined to be
$\tau_{sTH}=521.0579$.
The graphs of Fig.~(\ref{vi}) show the time dependence of the
master system state
variables (solid lines) and the corresponding slave system
state variables (dashed lines).
 From the graphs in Fig.~(\ref{eu}) we confirm that the
synchronization is reached in finite-time.
It is observed that synchronization is reached at
time $\tau_{sNU}\simeq 245.3$ which respects the finite-time condition
$\tau_{sNU}\leq \tau_{sTH}$ (See Fig.~(\ref{eu}b).\\

\begin{figure}[htp]
 \begin{minipage}[b]{8cm}
      \begin{center}
    \includegraphics[scale=0.16]{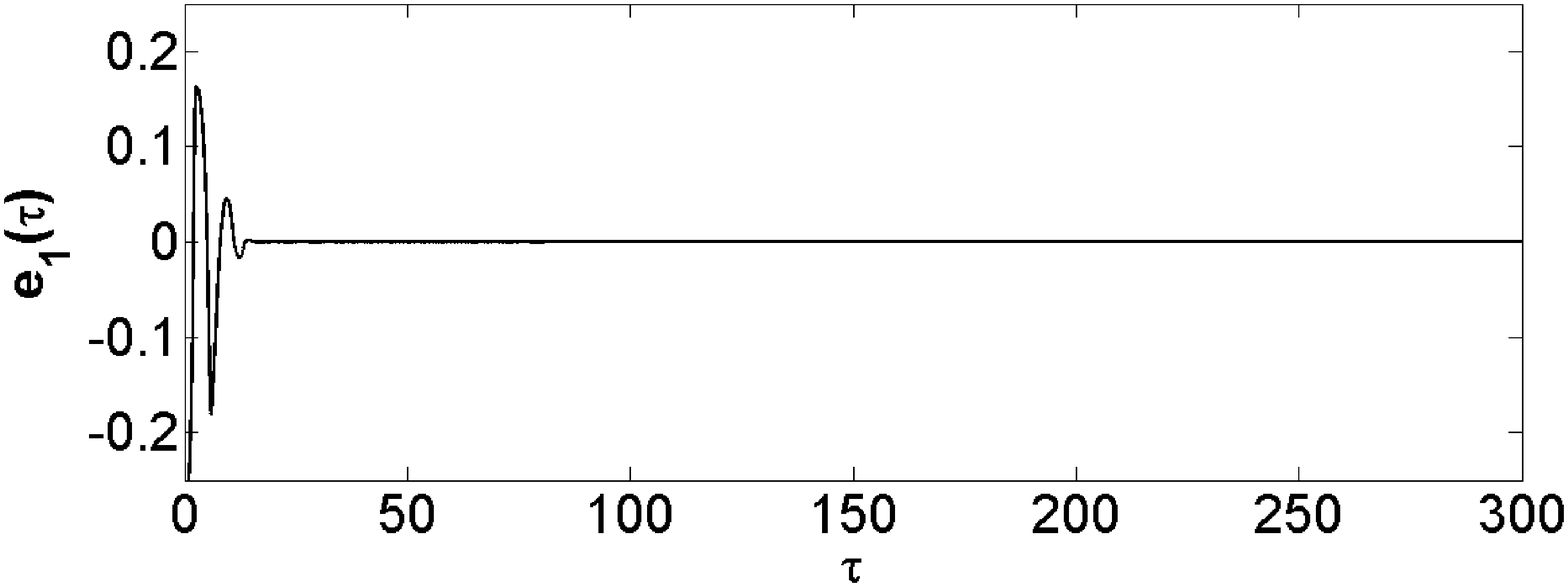}(a)
        \includegraphics[scale=0.16]{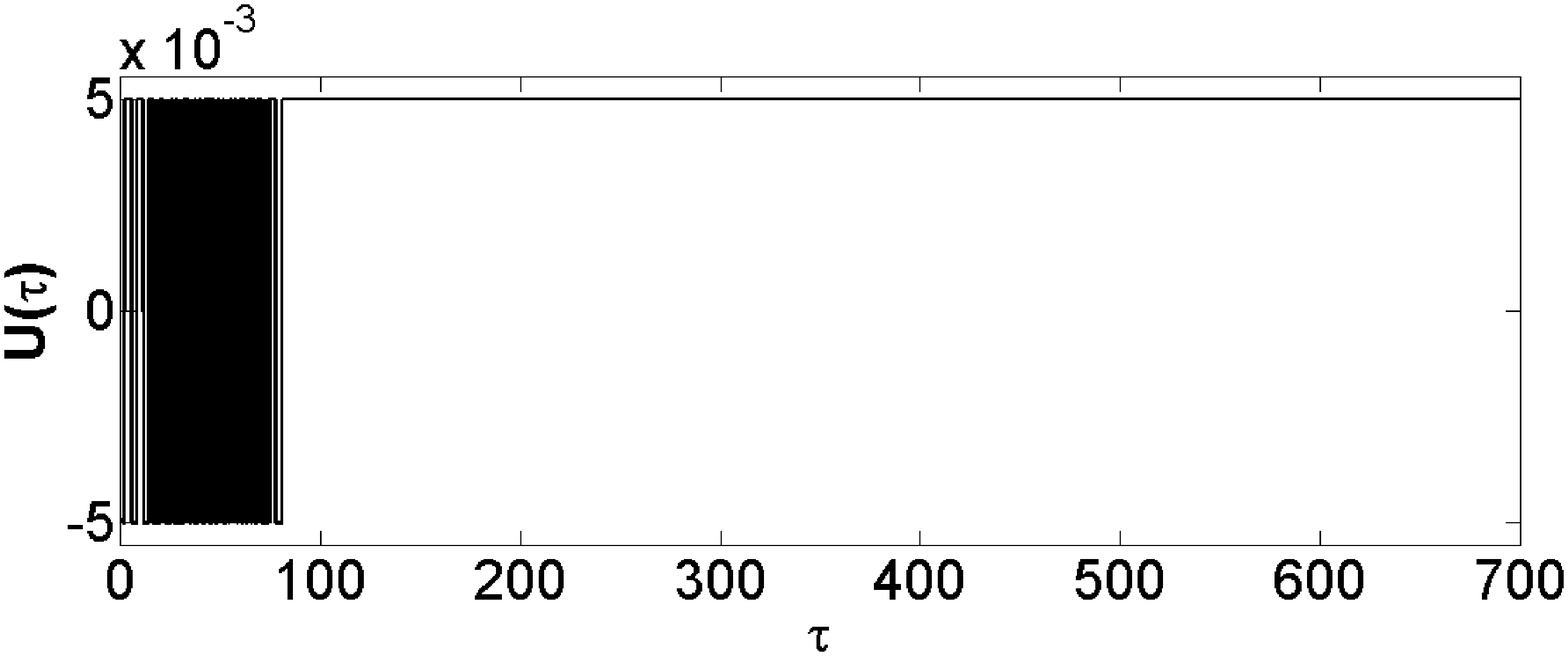}(b)
     \includegraphics[scale=0.16]{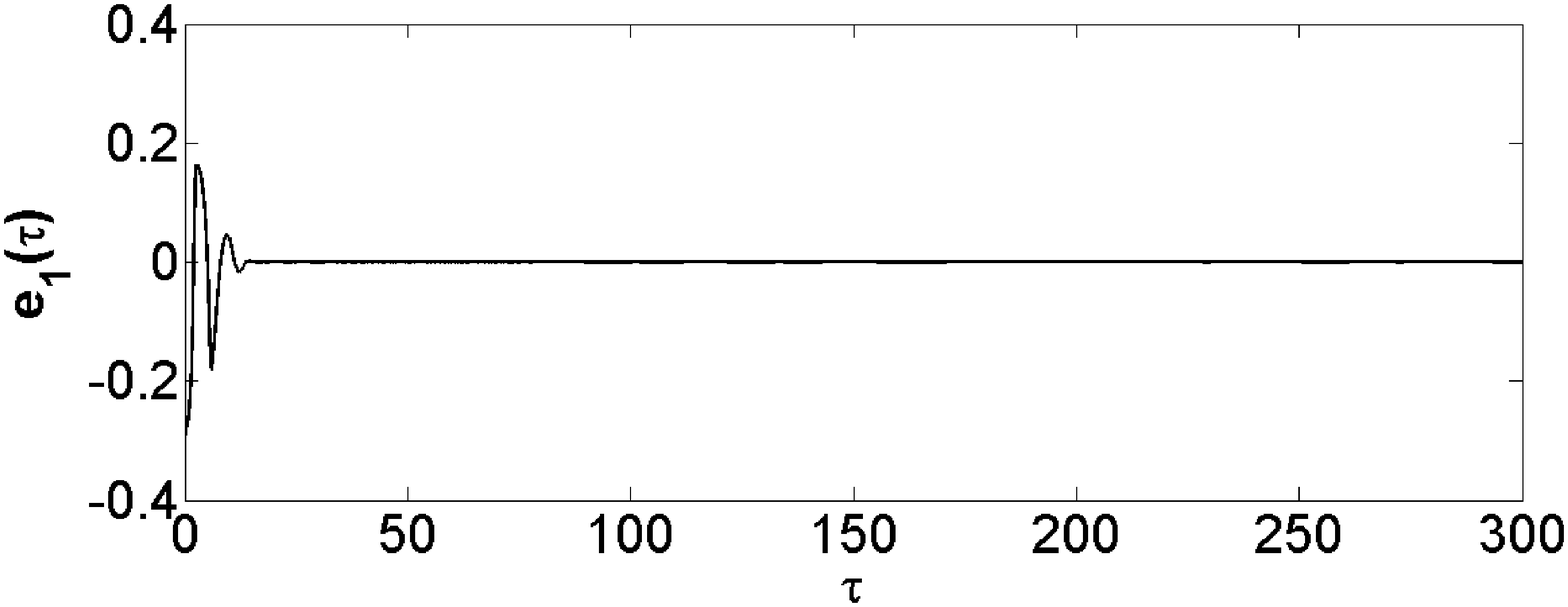}(c)
        \includegraphics[scale=0.16]{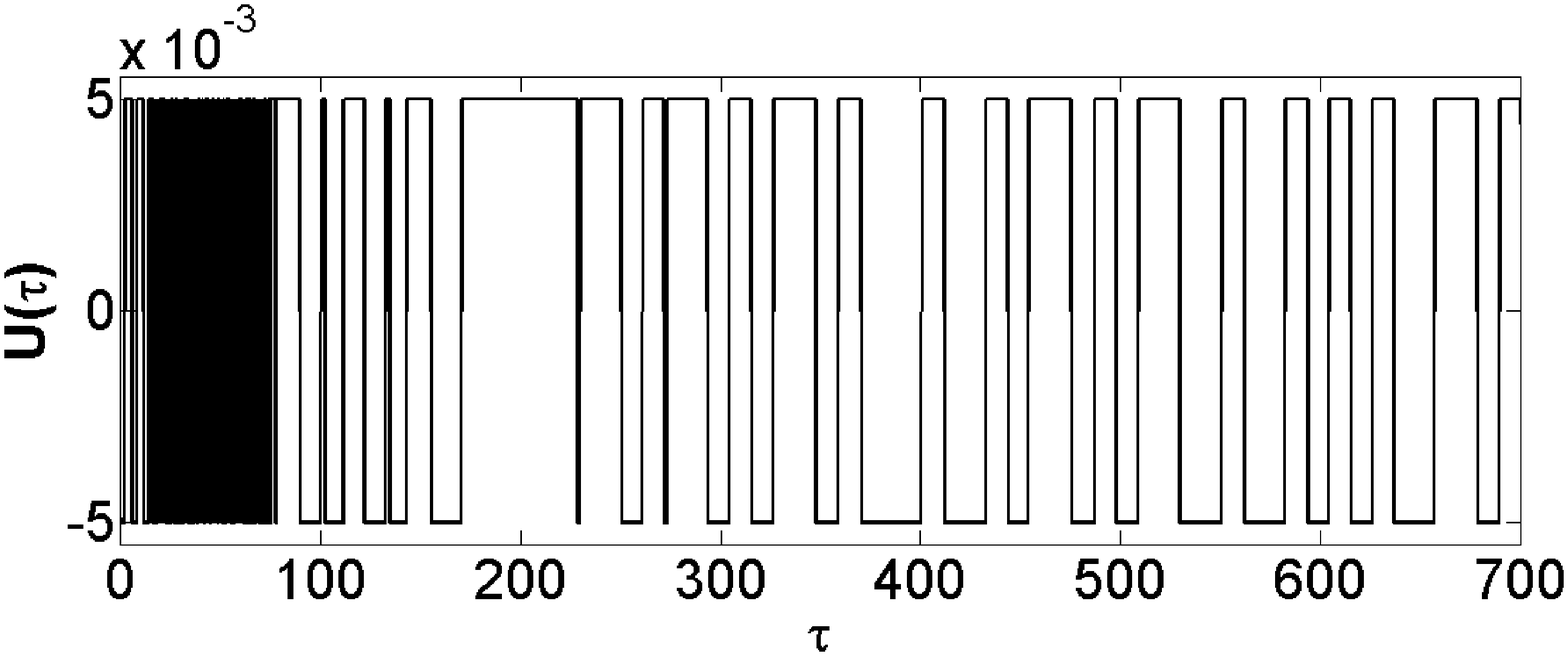}(d)
    \caption{Behaviors of $e_1(\tau)$ and $u(\tau)$. Figs.(a) and (b):
$p=0.0017$. Figs.(c) and (d): $p=0.02$.}
     \label{eip}
     \end{center}
     \end{minipage}
     \end{figure}

The value of the finite-time of stability depends strongly on the
values of parameters $p$. Considering $k\simeq 0.005$
the following graphs represent the behavior of the error state $e_1(\tau)$ and
of the adaptive parameter $u(\tau)$ for two given values
of $p$ (Fig.~(\ref{eip}.a)). In principle $p$ can take any values, but to deal
with finite-time stability, it
should be small.
The graphs in Fig.~(\ref{eip}) represent the behavior of both $e_1(\tau)$ and
$u(\tau)$ for $p=0.0017$ and $p=0.002$ (See Figs.~(\ref{eip})(a) and (b)
and Figs.~(\ref{eip})(c) and (d) respectively). In this paper we define the
numerical finite-time of convergence as the end time of the
transitory phase of
the time evolution of $e_1(\tau)$. For the graphs on Fig.~(\ref{eip}a) and
Fig.~(\ref{eip}b) the numerical finite-time is $\tau_{sNU} \simeq 100$ and the
theoretical finite-time is $\tau_{sTH}\simeq 306.5047$ while for the ones on
Fig.~(\ref{eip}c)
and Fig.~(\ref{eip}d) the theoretical finite-time is $\tau_{sTH}\simeq
260.5290$ and the numerical finite-time $\tau_{sNU} > \tau_{sTH}$.
Thus we see that the finite-time condition
$\tau_{sNU}\leq \tau_{sTH}$ is fulfilled in both cases.\\

\newpage
\section{Finite-time time-delay synchronization of CDT}
\subsection{Systems with internal delay}
\subsubsection{Synchronization analysis}
\noindent

In this section, we investigate the finite-time synchronization of delayed
tunnel diode based chaotic oscillators.
We first consider the presence of one delay affecting the nonlinearity of each
drive system and response system. Thus,
the master and slave systems become respectively
\begin{equation}
 \label{idd}
\left\{\begin{array}{lcl}
\dot{x}_1(\tau)&=&\alpha [x_2-x_1-rf(x_1(\tau-\theta))],\\
\\
\dot{x}_2(\tau)&=&\beta(x_1-x_2+ r x_3),\\
\\
\dot{x}_3(\tau)&=&\gamma(E-x_2),
       \end{array}
\right.
\end{equation}
and
\begin{equation}
\label{idr} \left\{ \begin{array}{lcl}
\dot{y}_1&=&\alpha \left[y_2-y_1-r f(y_1(\tau-\theta))\right]\\
&-&\zeta k~sign(y_1-x_1)-\zeta u(\tau),\\
\\
\dot{y}_2&=&\beta\left[y_1-y_2+ r y_3-2(y_1-x_1)\right],\\
\\
\dot{y}_3&=&\gamma \left[E-y_2\right],\end{array} \right.
\end{equation}
where $\theta$ is the time-delay.\\
For such a case, we define the functions
$\phi_1(x_1(\tau-\theta),
y_1(\tau-\theta))=r\left(f(y_1(\tau-\theta))-f(x_1(\tau-\theta))\right)$ and
$e_1(\tau-\theta)=y_1(\tau-\theta)-x_1(\tau-\theta)$. Thus the error state is
given by the following set of equations

\begin{equation}\label{ided} \left\{ \begin{array}{lcl}
\dot{e}_1&=&\alpha
\left[e_2-e_1-\phi_1(x_1(\tau-\theta),y_1(\tau-\theta))\right]\\
&-&\zeta k sign(e_1)-\zeta u(\tau),\\
\\
\dot{e}_2&=&\beta\left[-e_1-e_2+r e_3\right],\\
\\
\dot{e}_3&=&-\gamma e_2,\end{array} \right.
\end{equation}

From here, using the same controller as in the previous section, we choose the
following Krasovskii-Lyapunov function candidate~\cite{lia,sen}:

\begin{equation}\label{idly}\begin{array}{lcl}
V
&=&\displaystyle\frac{1}{2}\left(\displaystyle\frac{e_1^2}{\alpha}
+\displaystyle\frac{e_2^2}{\beta}
+\displaystyle\frac{r e_3^2}{\gamma}\right)+| u(\tau)|\\
&+& \eta \int_{-\theta}^{0}\!  e_1^{2} \left( \tau+s \right){ds},
\end{array}\end{equation}
where $\eta$ is a positive constant to be determined.\\
The time derivative along the trajectories of the system Eq.~(\ref{ided})
yields

\begin{equation}
\label{didl1} \begin{array}{lcl}
\dot{V} &=&-(1-\eta)e_1^2-e_2^2-\phi_1(\tau-\theta) e_1-\displaystyle\frac{\zeta
k}{\alpha} |e_1|\\
&-&\displaystyle\frac{\zeta u(\tau) }{\alpha}e_1+sign(u)\dot{u}(\tau)- \eta
e_1^2(\tau-\theta),\\
\\
&\leq& -(1-\eta)e_1^2-e_2^2+|\phi_1(\tau-\theta)||e_1| -
\displaystyle\frac{\zeta k}{\alpha} |e_1|\\
&-&\displaystyle\frac{\zeta u(\tau)
}{\alpha} e_1+sign(u)\dot{u}(\tau)- \eta e_1^2(\tau-\theta).
\end{array}
\end{equation}
Let assume that $|\phi_1(\tau-\theta)|\leq
\chi_1|e_1(\tau-\theta)|$, where
$\chi_1$ is a positive constant. It follows from here that

\begin{equation}
\label{didl2} \begin{array}{lcl}
\dot{V}&\leq& -e_1^2-e_2^2+\chi_1|e_1(\tau-\theta)||e_1| -
\displaystyle\frac{\zeta
k}{\alpha} |e_1|-\displaystyle\frac{\zeta u(\tau) }{\alpha} e_1\\
&+&sign(u)\dot{u}(\tau)+\eta e_1^2- \eta e_1^2(\tau-\theta),\\
\\
\dot{V}&\leq& -\left(1-\displaystyle\frac{\chi_1}{2}-\eta
\right)e_1^2-e_2^2+\left(\displaystyle\frac{\chi_1}{2}-\eta
\right)e_1^2(\tau-\theta)\\
&-& \displaystyle\frac{\zeta k}{\alpha}
|e_1|-\displaystyle\frac{\zeta u(\tau) }{\alpha} e_1+sign(u)\dot{u}(\tau).
\end{array}
\end{equation}

Let $\eta = \displaystyle\frac{\chi_1}{2}$, $\chi_1 < 1$ and $\zeta = \alpha$,
it follows that
\begin{equation}
\label{didl3} \begin{array}{lcl}
\dot{V} &\leq& - k |e_1|- u(\tau) e_1 +sign(u)\dot{u}(\tau).
\end{array}
\end{equation}

Thus, using the following controller

\begin{equation}\label{cont1}
\dot{u}= sign(u)\left(k|e_1|+ u e_1- p \right),
\end{equation}
it follows that

\begin{equation}
\label{didl4} \begin{array}{lcl}
\dot{V} &\leq& -p.
\end{array}
\end{equation}

Hence, global finite-time stability is achieved~\cite{zup}. For any time $\tau$
contained in the interval $0 < \tau < \theta$ both the drive and response system
do not oscillate at the considered regime. Therefore, to determine the
theoretical
finite settling time, we integrate Eq.~(\ref{didl4}) from $\theta$ to
$\tau_s$ and we obtain:

\begin{equation}
\label{idli}
V(\tau_s)-V(\theta) \leq -p \left(\tau_s-\theta \right).
\end{equation}
Using  the fact that  $V(\tau_s)\rightarrow 0$ when $\tau \rightarrow \tau_s$,
$V > 0~\forall~\tau$ and taking into account   Eq.~(\ref{didl4}),
we can see that the Lyapunov function $V$ is a monotonous and decreasing
function, it follows that

\begin{equation} \label{idlts}\begin{array}{lcl}
\tau_s &=& \theta +
\displaystyle\frac{1}{2p}\left(\displaystyle\frac{1}{\alpha}
e_1^2(\theta)+\displaystyle\frac{1}{\beta}e_2^2(\theta)+
\displaystyle\frac{R}{\gamma}e_3^2(\theta)\right)+\displaystyle\frac{|u(\theta)|
}{p}\\
&+& \eta \int _{-\theta}^{0}\!  \varepsilon_{{1}}^{2} \left( \theta+s
\right){ds},\\
\\
&\geq& \theta +
\displaystyle\frac{1}{2p}\left(\displaystyle\frac{1}{\alpha}
e_1^2(\theta)+\displaystyle\frac{1}{\beta}e_2^2(\theta)+
\displaystyle\frac{R}{\gamma}e_3^2(\theta)\right)+\displaystyle\frac{|u(\theta)|
}{p}.
\end{array}\end{equation}

Eq.~(\ref{idlts}) gives the maximum settling time for synchronization. Hence,
the finite-time synchronization is reached  when the numerical settling time
$\tau_{sNU}$ satisfies  the relation $\tau_{sNU} < \tau_{sTH}$, where
$\tau_{sTH}=\theta +
\displaystyle\frac{1}{2p}\left(\displaystyle\frac{1}{\alpha}
e_1^2(\theta)+\displaystyle\frac{1}{\beta}e_2^2(\theta)+
\displaystyle\frac{R}{\gamma}e_3^2(\theta)\right)+\displaystyle\frac{|u(\theta)|
}{p}$.

\subsubsection{Numerical results}
\noindent

\begin{figure}[htp]
 \begin{minipage}[b]{8cm}
      \begin{center}
    \includegraphics[scale=0.18]{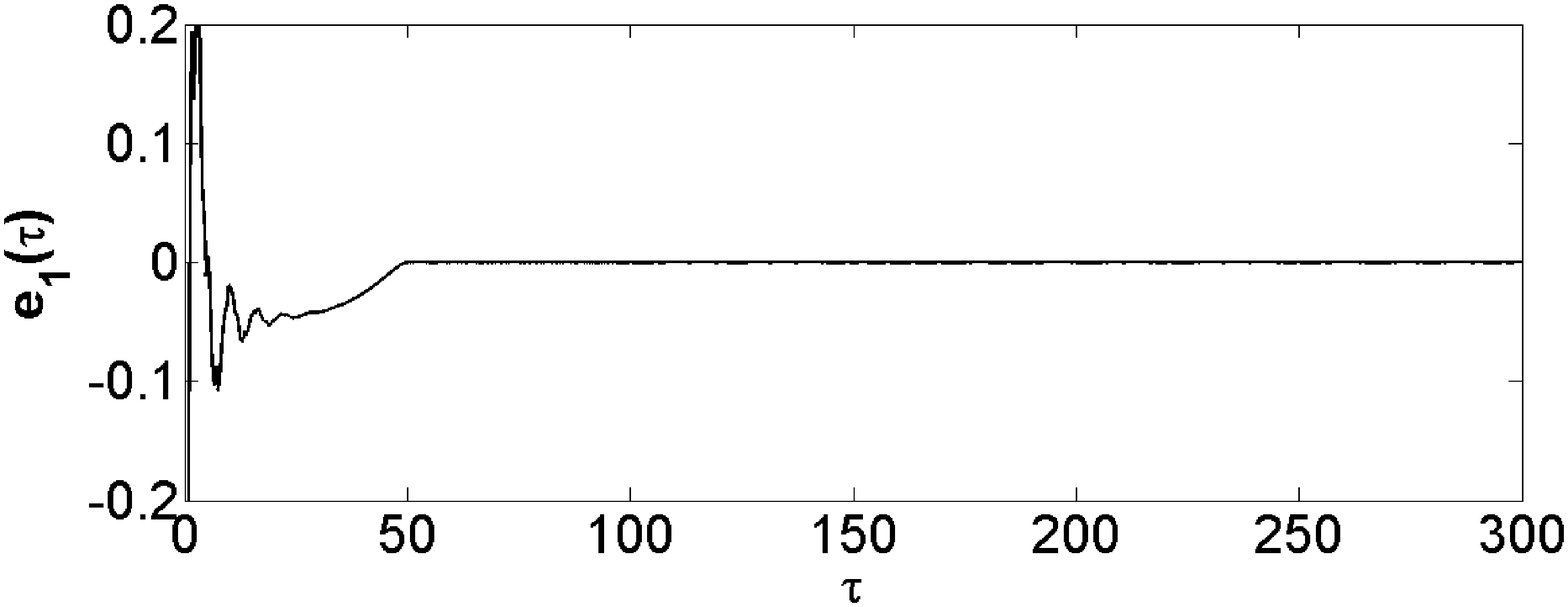}(a)
        \includegraphics[scale=0.18]{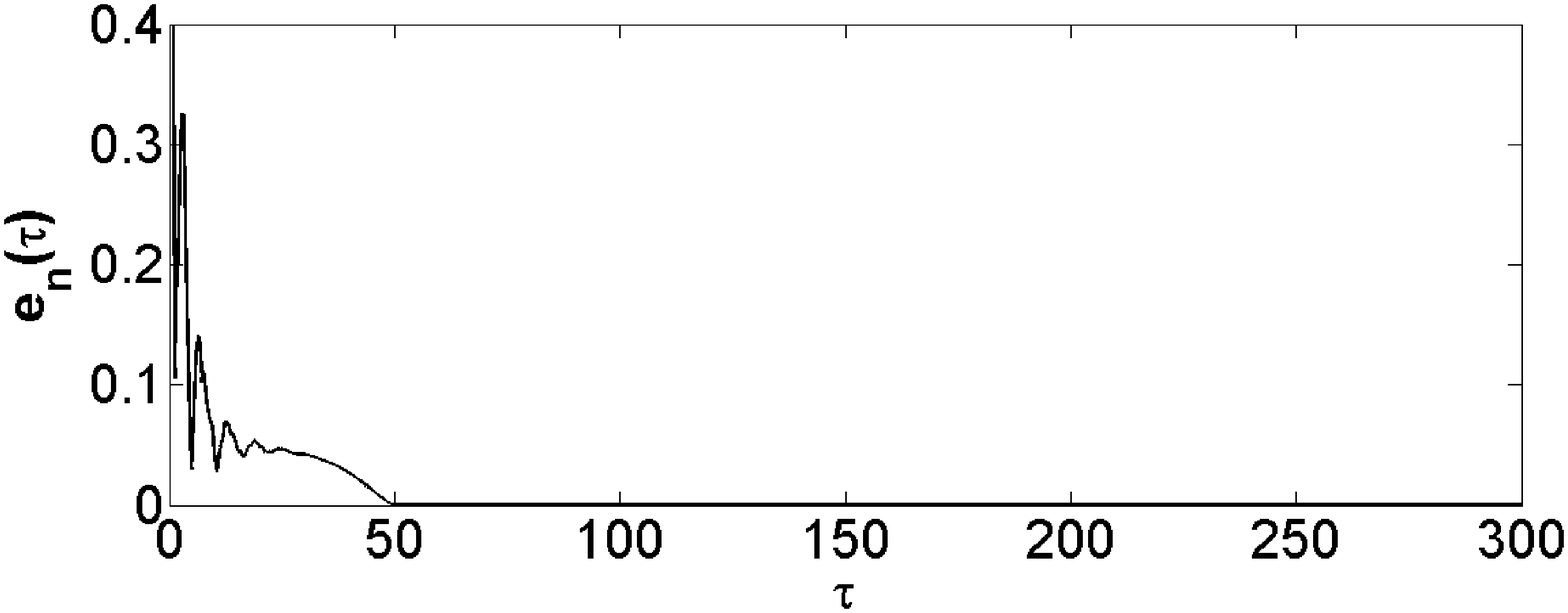}(b)
    \caption{Time history of $e_1(\tau)$ (a) and
$e_n(\tau)=\sqrt{e_1^2(\tau)+e_2^2(\tau)+e_3^2(\tau)}$ (b) .}
     \label{ider}
     \end{center}
     \end{minipage}
     \end{figure}

In this section we investigate the finite-time synchronization of two delayed
tunnel diode based chaotic systems basing
ourselves on the obtained numerical results of the established theory in the
previous subsection. The integration time step taken was $10^{-4}$.
Fig.~(\ref{ider}) helps to confirm the synchronization
behavior of both delayed master system (Eq.~(\ref{idd})) and slave system
(Eq.~(\ref{idr})) when $p=0.001$ and
$k=0.005$. With this selected values of parameters $p$ and $k$, the theoretical
settling time results $\tau_{sTH}\simeq 127.5448$,
while the numerical settling time corresponding to the condition given before is
$\tau_{sNU}\simeq 99.4$.
\newpage
When varying the time-lag $\theta$, just the theoretical settling time is
modified when synchronization occurs and the numerical finite-time is
determined as $\tau_{sNU}\simeq 99.4$.

\subsection{Multi-delayed systems}
\subsubsection{Synchronization analysis}
\noindent

In this section we consider that there exist two delays: $\theta_1$
represents the time-lag taken for the introduction of the nonlinearity and
$\theta_2$ being the time-delay between the master state and the
slave state. Thus, the systems become:
\begin{equation}
 \label{mdd}
\left\{\begin{array}{lcl}
\dot{x}_1(\tau-\theta_2)&=&\alpha
[x_2(\tau-\theta_2)-x_1(\tau-\theta_2)\\
&-&rf(x_1(\tau-\theta_1-\theta_2)],\\
\\
\dot{x}_2(\tau-\theta_2)&=&\beta(x_1(\tau-\theta_2)-x_2(\tau-\theta_2)\\
&+& r x_3(\tau-\theta_2)),\\
\\
\dot{x}_3(\tau-\theta_2)&=&\gamma(E-x_2(\tau-\theta_2)),
       \end{array}
\right.
\end{equation}
and
\begin{equation}
\label{mdr} \left\{ \begin{array}{lcl}
\dot{y}_1(\tau)&=&\alpha \left[y_2(\tau)-y_1(\tau)-r
f(y_1(\tau-\theta_1))\right]\\
&-&\zeta k~sign(y_1(\tau)-x_1(\tau-\theta_2))-\zeta
u(\tau),\\
\\
\dot{y}_2(\tau)&=&\beta\left[y_1(\tau)-y_2(\tau)+ r
y_3-2(y_1(\tau)-x_1(\tau-\theta_2))\right],\\
\\
\dot{y}_3(\tau)&=&\gamma \left[E-y_2(\tau)\right].\end{array} \right.
\end{equation}
In this case, we define the functions
$\phi_2(\tau,\theta_1,
\theta_2)=r\left(f(y_1(\tau-\theta_1))-f(x_1(\tau-\theta_1-\theta_2))\right)$
and
$e_1(\tau-\theta_1-\theta_2)=y_1(\tau-\theta_1)-x_1(\tau-\theta_1-\theta_2)$.
Thus the error state is given by the following set of equations

\begin{equation}\label{mder2} \left\{ \begin{array}{lcl}
\dot{e}_1&=&\alpha \left[e_2-e_1-\phi_2(\tau,\theta_1,\theta_2)\right]\\
&-&\zeta k sign(e_1)-\zeta u(\tau),\\
\\
\dot{e}_2&=&\beta\left[-e_1-e_2+r e_3\right],\\
\\
\dot{e}_3&=&-\gamma e_2,\end{array} \right.
\end{equation}

Let us now select the Krasovskii-Lyapunov function as~\cite{lia,sen}:

\begin{equation}\label{mdly}\begin{array}{lcl}
V&=&\displaystyle\frac{1}{2}\left(\displaystyle\frac{e_1^2}{\alpha}
+\displaystyle\frac{e_2^2}{\beta}
+\displaystyle\frac{r e_3^2}{\gamma}\right)+| u(\tau)|\\
&+& \lambda \int_{-\theta_1}^{0}\!  e_{{1}}^{2} \left( \tau-\theta_2+s
\right){ds},
\end{array}\end{equation}
where $\lambda$ is a positive constant to be determined.\\
The time derivative along the trajectories of the system Eq.~(\ref{mder2})
yields

\begin{equation}
\label{mddl1} \begin{array}{lcl}
\dot{V} &=&-(1-\lambda)e_1^2-e_2^2-\phi_2(\tau,\theta_1,\theta_2)
e_1-\displaystyle\frac{\zeta k}{\alpha} |e_1|\\
&-&\displaystyle\frac{\zeta u(\tau)
}{\alpha}e_1+sign(u)\dot{u}(\tau)- \lambda e_1^2(\tau-\theta_1-\theta_2),\\
\\
&\leq& -(1-\lambda)e_1^2-e_2^2+|\phi_2(\tau,\theta_1,\theta_2)||e_1| -
\displaystyle\frac{\zeta k}{\alpha} |e_1|\\
&-&\displaystyle\frac{\zeta u(\tau)
}{\alpha} e_1 +sign(u)\dot{u}(\tau)- \lambda e_1^2(\tau-\theta_1-\theta_2).
\end{array}
\end{equation}
Let assume that $|\phi_2(\tau,\theta_1,\theta_2)|\leq
\chi_2|e_1(\tau-\theta_1-\theta_2)|$, where $\chi_2$ is positive constant. It
follows from here that

\begin{equation}
\label{mddl2} \begin{array}{lcl}
\dot{V}&\leq&
-e_1^2-e_2^2+\chi_2|e_1(\tau-\theta_1-\theta_2)||e_1| -
\displaystyle\frac{\zeta
k}{\alpha} |e_1|\\
&-&\displaystyle\frac{\zeta u(\tau) }{\alpha} e_1
+sign(u)\dot{u}(\tau)+\lambda e_1^2- \lambda e_1^2(\tau-\theta_1-\theta_2),\\
\\
\dot{V}&\leq& -\left(1-\displaystyle\frac{\chi_2}{2}-\lambda
\right)e_1^2+\left(\displaystyle\frac{\chi_2}{2}-\lambda
\right)e_1^2(\tau-\theta_1-\theta_2)\\
&-& \displaystyle\frac{\zeta k}{\alpha}
|e_1|-e_2^2-\displaystyle\frac{\zeta u(\tau) }{\alpha} e_1 +sign(u)\dot{u}(\tau).
\end{array}
\end{equation}

Let $\lambda = \displaystyle\frac{\chi_2}{2}$, $\chi_2 < 1$ and $\zeta =
\alpha$, it follows that
\begin{equation}
\label{mddl3} \begin{array}{lcl}
\dot{V} &\leq& - k |e_1|- u(\tau) e_1 +sign(u)\dot{u}(\tau).
\end{array}
\end{equation}

Thus, using the following controller Eq.~(\ref{cont2}),

\begin{equation}\label{cont2}
\dot{u}= sign(u)\left(k|e_1|+ u e_1- p \right),
\end{equation}
it follows that

\begin{equation}
\label{mddl4} \begin{array}{lcl}
\dot{V} &\leq& -p.
\end{array}
\end{equation}

Hence, the global finite-time stability is achieved~\cite{zup}. When $0 < \tau
< \theta_1+\theta_2$ the system can not oscillate at the considered regime.
In order to determine the theoretical finite settling time, we integrate the
Eq.~(\ref{mddl4}) from $\theta_1+\theta_2$ to $\tau_s$. Thus, the finite
settling time is given by:

\begin{equation} \label{mdts}\begin{array}{lcl}
\tau_s &=&\displaystyle\frac{1}{2p}\left(\displaystyle\frac{1}{\alpha}
e_1^2(\theta_1+\theta_2)+\displaystyle\frac{1}{\beta}e_2^2(\theta_1+\theta_2)+
\displaystyle\frac{R}{\gamma}e_3^2(\theta_1+\theta_2)\right)\\
&+&\displaystyle\frac{|u(\theta_1+\theta_2)|}{p}
+ \lambda \int _{-\theta_1+\theta_2}^{0}\!  \epsilon_{{1}}^{2} \left(
\theta_1+\theta_2+s \right){ds}+\theta_1+\theta_2 ,\\
\\
&\geq&\displaystyle\frac{1}{2p}\left(\displaystyle\frac{1}{\alpha}
e_1^2(\theta_1+\theta_2)+\displaystyle\frac{1}{\beta}e_2^2(\theta_1+\theta_2)+
\displaystyle\frac{R}{\gamma}e_3^2(\theta_1+\theta_2)\right)\\
&+&\displaystyle\frac{|u(\theta_1+\theta_2)|}{p}+\theta_1+\theta_2 .
\end{array}\end{equation}
At this moment, as in the previous section, the finite-time synchronization is
reached  when the numerical settling time $\tau_{sNU}$ satisfies  the relation
$\tau_{sNU} < \tau_{sTH}$, where
\begin{displaymath}\begin{array}{lcl}
\tau_{sTH}&=&\displaystyle\frac{1}{2p}\left(\displaystyle\frac{1}{\alpha}
e_1^2(\theta_1+\theta_2)+\displaystyle\frac{1}{\beta}e_2^2(\theta_1+\theta_2)+
\displaystyle\frac{R}{\gamma}e_3^2(\theta_1+\theta_2)\right)\\
&+&\displaystyle\frac{
|u(\theta_1+\theta_2)|}{p}+\theta_1+\theta_2.
\end{array}\end{displaymath}

\subsubsection{Numerical results}
\noindent

\begin{figure}[htp]
 \begin{minipage}[b]{8cm}
      \begin{center}
    \includegraphics[scale=0.18]{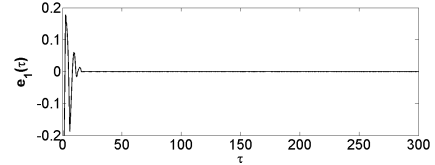}
   \caption{Time history of $e_1(\tau)$.}
     \label{mder}
     \end{center}
     \end{minipage}
     \end{figure}

\begin{figure}[htp]
 \begin{minipage}[b]{8cm}
      \begin{center}
    \includegraphics[scale=0.12]{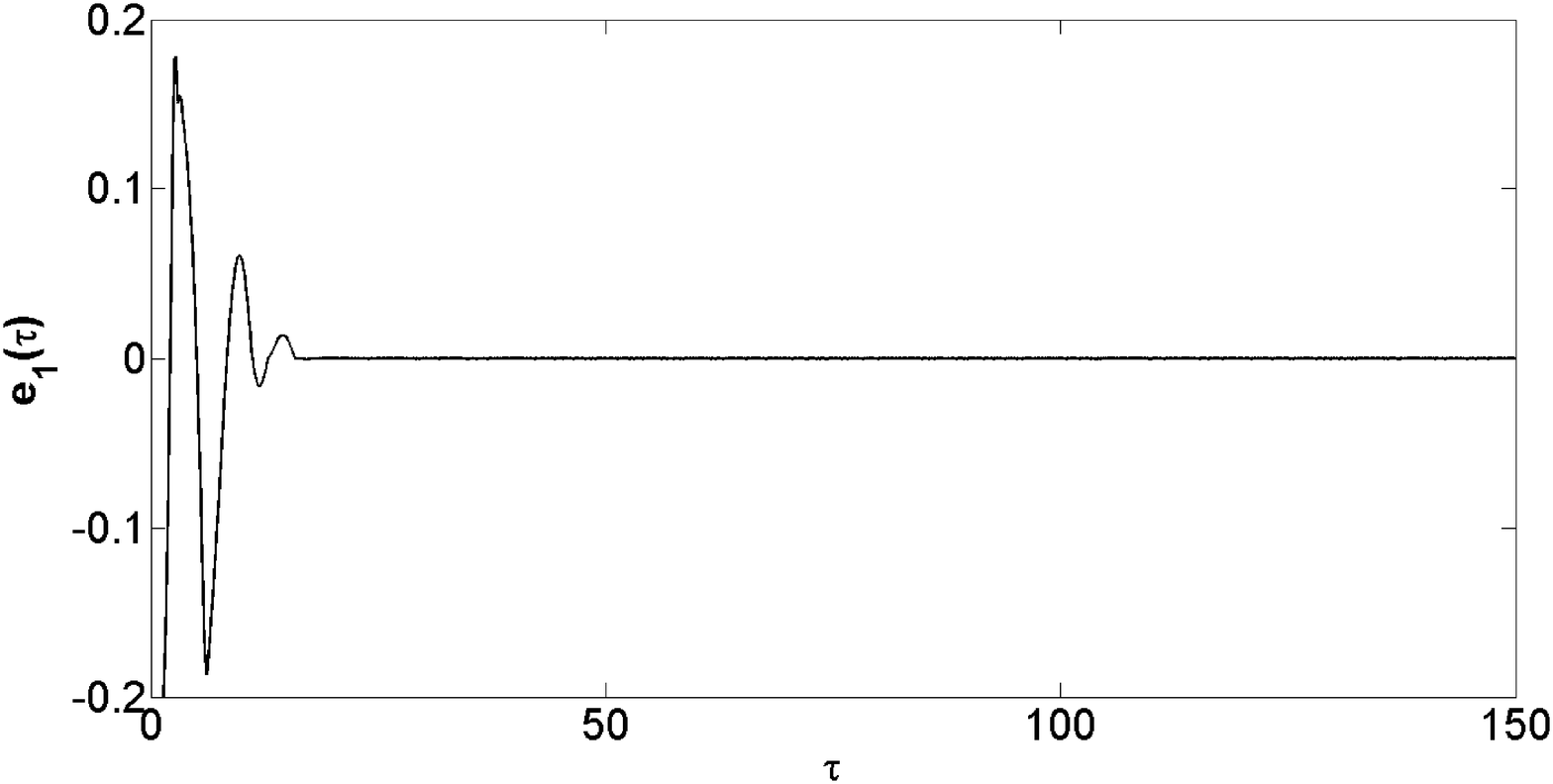}(a)
        \includegraphics[scale=0.13]{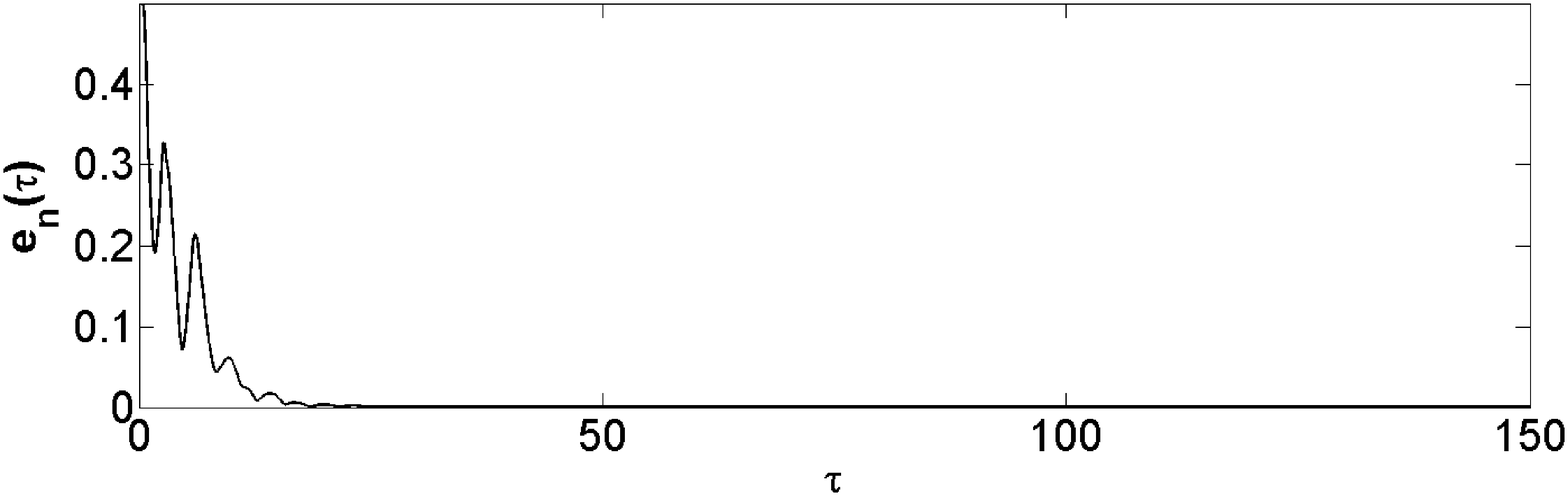}(b)
   \caption{Time history of (a) $e_1(\tau)$ and (b) $e_n(\tau)$ when
$\theta_2=0.0003$.}
     \label{mdei1}
     \end{center}
     \end{minipage}
     \end{figure}

\begin{figure}[htp]
 \begin{minipage}[b]{8cm}
      \begin{center}
            \includegraphics[scale=0.14]{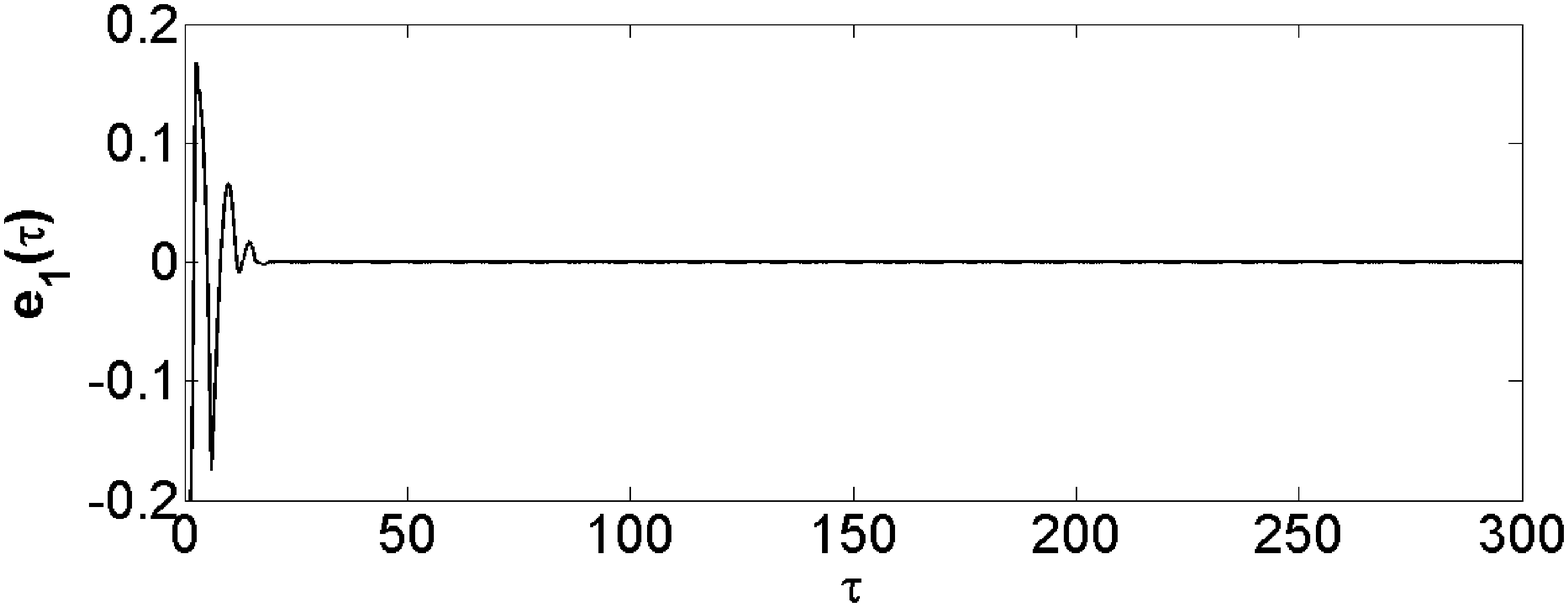}(a)
        \includegraphics[scale=0.14]{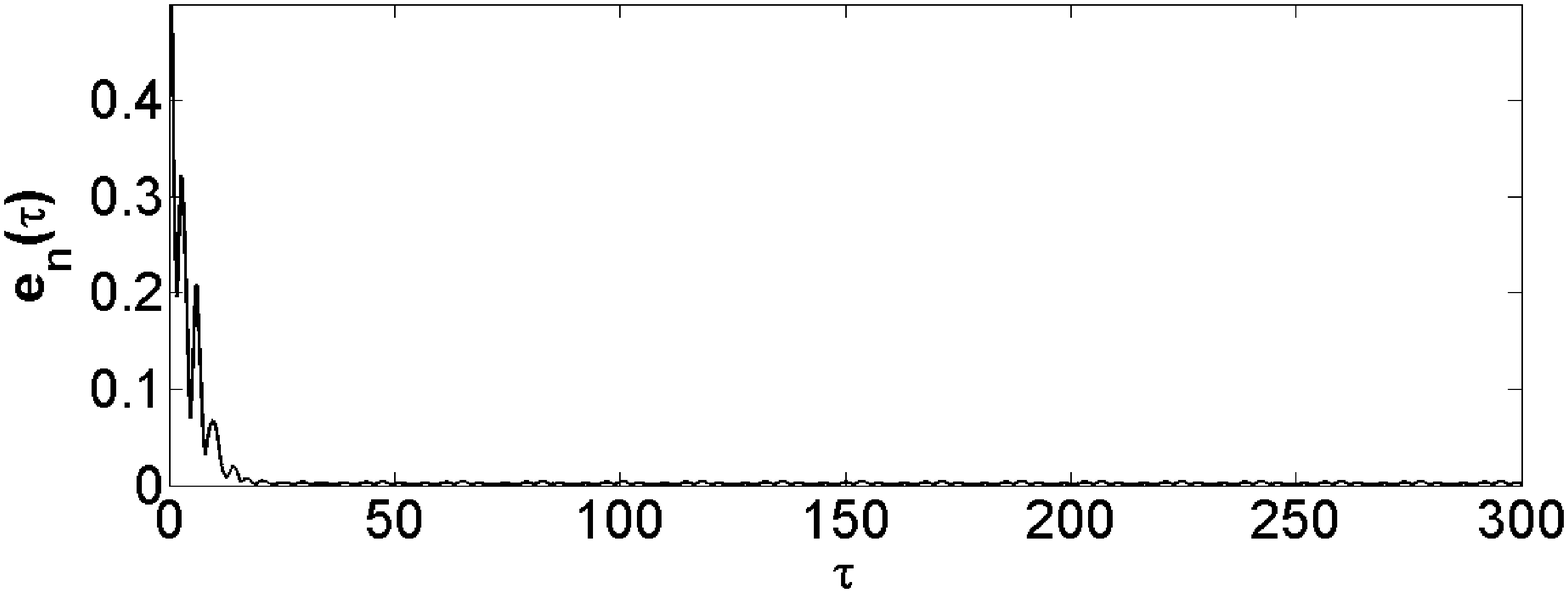}(b)
   \caption{Time history of (a) $e_1(\tau)$ and (b) $e_n(\tau)$ when
$\theta_2=0.004$.}
     \label{mdei2}
     \end{center}
     \end{minipage}
     \end{figure}

\begin{figure}[tp]
 \begin{minipage}[b]{8cm}
      \begin{center}
            \includegraphics[scale=0.14]{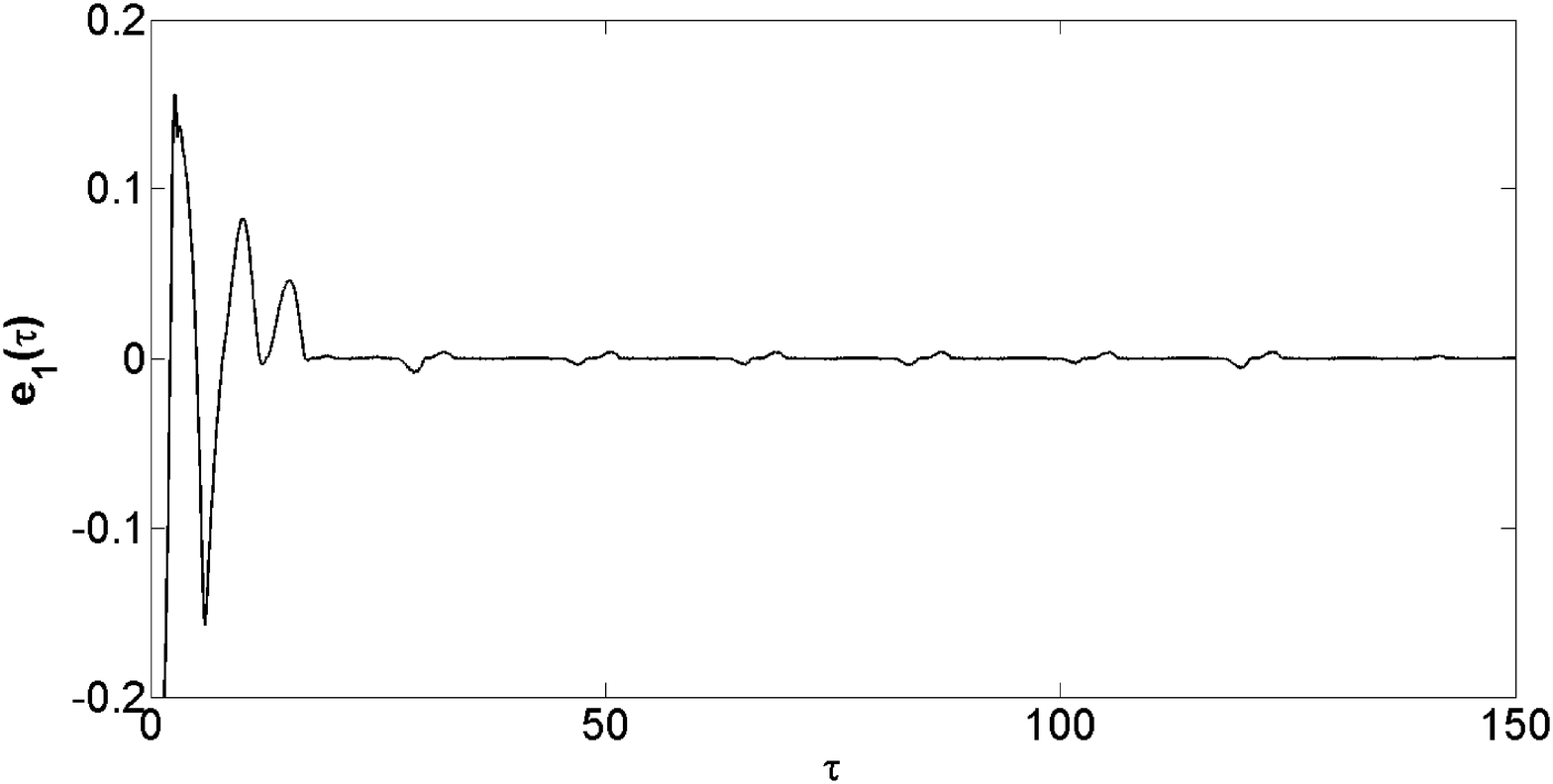}(a)
        \includegraphics[scale=0.14]{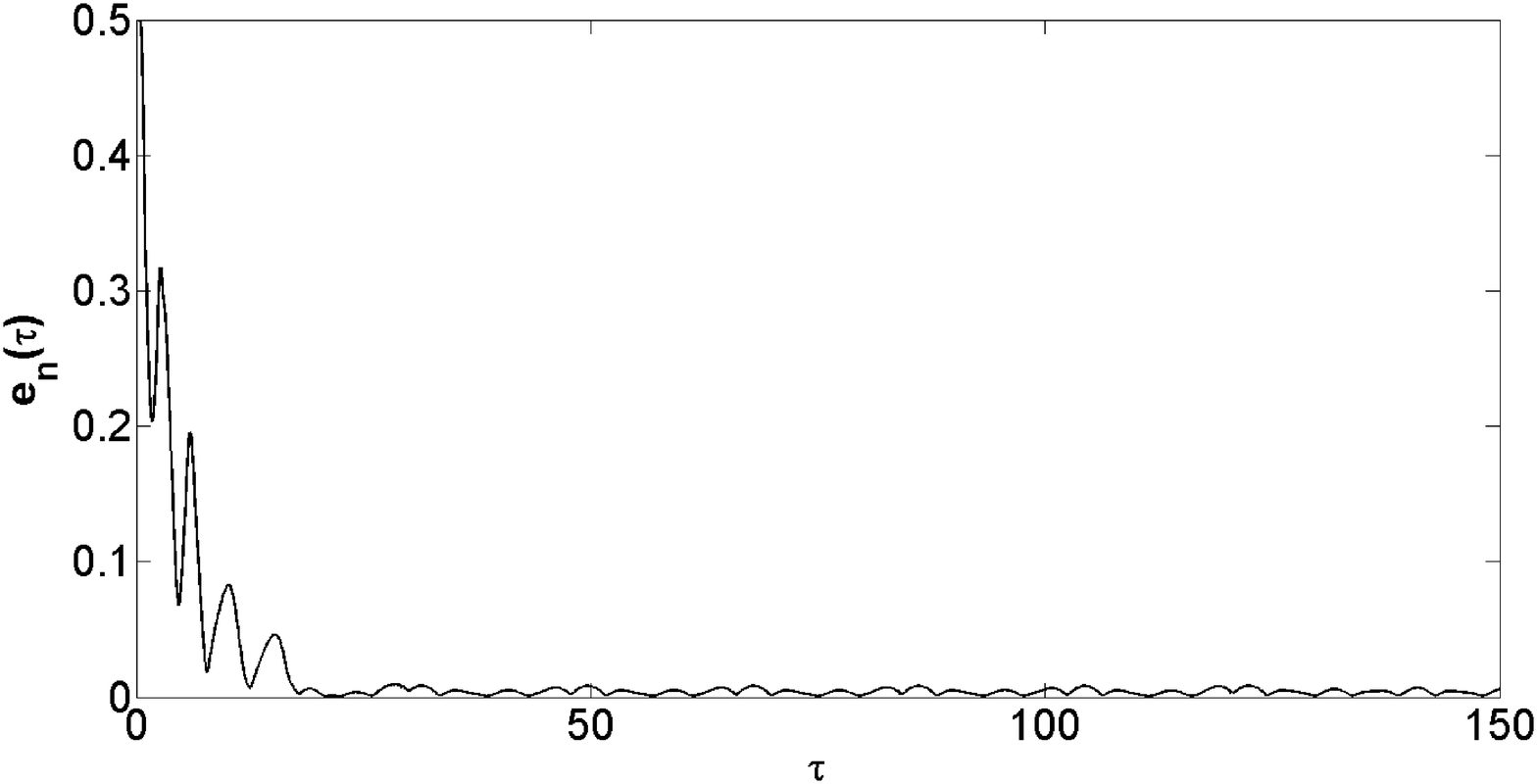}(b)
   \caption{Time history of (a) $e_1(\tau)$ and (b) $e_n(\tau)$ when
$\theta_2=0.08$.}
     \label{mdei3}
     \end{center}
     \end{minipage}
     \end{figure}

From our investigations, it comes out that, the finite-time synchronization is
hardly reached when we consider that both dynamics
of the drive system and response system are subjected to delay $\theta_2$. The
integration time step used in this part was $10^{-4}$.
The following graphs on Fig.~(\ref{mder}) show that the goal could be
achieved
only if $\theta_2$ is relatively small. For the following values
$k=0.005$, $p=0.001$, $\theta_1=0.1$ and $\theta_2=0.0003$, one has the
following behavior.\\
Considering the behavior of the synchronization error $e_1(\tau)$ and
$e_n(\tau)$ (defined in Fig.~(\ref{ider})), one can observe the destruction of
the
synchronization when we increase the value of $\theta_2$ (See
Figs.~(\ref{mdei1}), Figs.~(\ref{mdei2}) and Figs.~(\ref{mdei3})).

\newpage
\section{Conclusions}
The target of this paper was to investigate the possibility to achieve
synchronization in finite-time of
two tunnel diode based chaotic oscillators. We have considered the case of
chaotic systems without and with delay (internal delay and multiple delays). The
controller was built basing ourselves on the absolute stability
theory~\cite{lia} and
on the Krasovskii-Lyapunov stability theory~\cite{sen}. Later, the expression of
the
settling finite-time was investigated in all considered cases. Numerical
simulations were performed and given to confirm our theoretical analysis. We
observe that the finite-time synchronization was reached when some conditions,
which are given and proved by numerical results, are filled.

\bibliography{sample-paper}

\begin{thebibliography}{10}


\bibitem{bla} B. Blasius,  A. Huppert and  L. Stone, Nature  \textbf{399}, 354
(1999).

\bibitem{gra} D. W. Graham, C. W Knapp, E. S Van Vleck, K. Bloor, Teresa
B. Lane and C. E. Graham, Int. Soc. Microb. Ecol. J. \textbf{1}, 385 (2007).

\bibitem{lou} P. Louodop, H. Fotsin and S. Bowong,  Phys. Scr. \textbf{85},
025002 (2012).

\bibitem{bow1} S. Bowong and J. Kurths, Phys. Lett. A \textbf{374}, 4496 (2010).

\bibitem{casa}A. Casadevall,f. C. Fang, L-a. Pirofski, PLoS Pathog \textbf{7}
1002136 (2011).


\bibitem{phil} P. Philippe, Hum. Biol. \textbf{65}, 525 (1993).

\bibitem{earn} D. J. D. Earn, P. Rohanian and B. Y. Grenfell, Proc R. Soc. Lond.
B \textbf{265}, 7 (1998).

\bibitem{benhu} E. Beninca, J. Huisman, R Heerkloss, K. D. Johnk, P. Branco, E.
H. van Nes, M. Scheffer and S. P. Ellner, Nature \textbf{451}, 822 (2008).

\bibitem{cru} C. Cruz-Hernandez, Nonlinear Dynamics and Systems Theory
\textbf{1}, 1 (2004).

\bibitem{bow2} S. Bowong and J.J. Tewa, Math. Comput. Model.  \textbf{48}, 1826
(2008).

\bibitem{fek} M. Feki,     Chaos Solitons Fractals  \textbf{18}, 141 (2003).

\bibitem{yan} W. Yang, X. Xia, Y. Dong and S. Zheng, Computer and Information
Science \textbf{3} 174 (2010).

\bibitem{gui} Z. Gui, X. Wu and Q. Lin, in \emph{Proceedings of the 2009
International Workshop on Chaos-Fractals Theories and Applications},
2009, (Ieee Computer Society Washington, DC, USA, 2009),p.16.

\bibitem{laz} M. P. Lazareviæ and D. Lj. Debeljkoviæ, Asian J. Control
\textbf{7}, 440 (2005).

\bibitem{zup} L. A. Zuppa, C. C. H\'ernandez and A. Y. A.
Bustos, Finite synchronization of Lorenz-based chaotic systems, 2002,
http://www.wseas.us/e-library/conferences/mexico2002/papers/249.pdf.

\bibitem{lak} M. Lakshmanan, D. V. Senthilkumar, \emph{Dynamics of Nonlinear
time-delay systems}(Springer-verlag, New York, 2010).

\bibitem{bro} E. R. Brown, J. R. Soderstrom, C. D. Parker, L. J.
Mahoney, K. M. Molvar and T. C. McGill, Appl. Phys. Lett.  \textbf{58}, 2291
(1991).

\bibitem{suz} S. Suzuki, A. Teranishi, K. Hinata, M. Asada, H.
Sugiyama and H. Yokoyama, Appl. Phys. Express.  \textbf{58}, 192 (2009).

\bibitem{liq} L. Wang, \emph{Reliable design of tunnel diode and
resonant
tunnelling diode based microwave sources. PhD thesis}(Glasgow University, 2011,
http://theses.gla.ac.uk/3423/).

\bibitem{wir} H. Ren, M. S. Baptista and C. Grebogi, Physical
Review Letters \textbf{110}, 184101-1 (2013).

\bibitem{fir} B. S. Dmitriev, A. E. Hramov, A. A. Koronovskii,
A. V. Starodubov,
D. I. Trubetskov and Y. D. Zharkov, Physical Review Letters \textbf{102}
074101-1 (2009).

\bibitem{sec} A. A. Koronovskii, O. I. Moskalenko, A. E. Hramov,
Physics-Uspekhi  \textbf{52}, 1213 (2009).

\bibitem{yon} Y. Liu, Nonlinear Dynamics  \textbf{67}, 89
(2012).

\bibitem{hua} H. Wang, Z. Han, Q. Xie, W. Zhang, Nonlinear
Analysis: Real World Applications \textbf{10}, 2842 (2009).


\bibitem{pik} A. S. Pikovskii and M. I. Rabinovich, Sov. Phys. Dokl.
\textbf{23}, 183 (1978).

\bibitem{mar} A. Y. Markov, A. L. Fradkov and G. S. Simin, in \emph{Proceeding
of 35th Conference on Decision and Control}, 1996,(Kobe, 1996),p.2177.

\bibitem{yas} M. T. Yassen, Appl. Math. Comput. \textbf{135}, 113 (2003).

\bibitem{lia} X. Liao and P. Yu, \emph{Absolute stability of nonlinear control
systems}(Springer science + Buisness Media B. V, New York, 2008).

\bibitem{spr} J. C. Sprott, Phys. Lett. A \textbf{266}, 19 (2000).

\bibitem{lin} J. Lin, C. Huang, T. Liao and J. Yan, Digit. Signal Process.
\textbf{20}, 229 (2010).

\bibitem{sen} D. V. Senthilkumar, J. Kurths and M. Lakshmanan, Phys. Rev. E
\textbf{79}, 066208 (2009).

\end{thebibliography}

\bibliographystyle{prsty}

\section*{Appendx: Circuit design and simulations}
\noindent

\begin{figure}[htp]
 \begin{minipage}[b]{8cm}
      \begin{center}
        \includegraphics[scale=0.4]{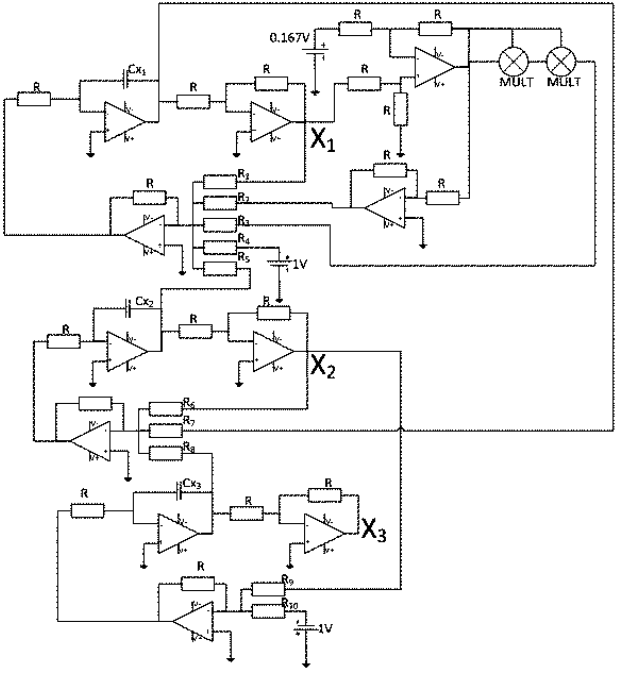}
   \caption{Circuit diagram of the analogue CDT system.}
     \label{anlc}
     \end{center}
     \end{minipage}
     \end{figure}

Here we investigate the electrical circuit of
the system. First we build  the analogue computer
of the chaotic oscillator, which is shown in Fig.~(\ref{anlc}). The use of the
analogue computer here is
helpful because we can modify the frequency of the oscillator
and thus facilitate the design of the controller.\\
%
%
This circuit is obtained through the following relations:

$\alpha =\displaystyle\frac{1}{\chi R_1C_{x_1}}=\displaystyle\frac{1}{\chi R_5
C_{x_1}}$, $\alpha r a_1 =\displaystyle\frac{1}{\chi R_3 C_{x_1}}$
, $\alpha r a_2 =\displaystyle\frac{1}{\chi R_2 C_{x_1}}$, $\alpha r a_3
=\displaystyle\frac{1}{\chi R_4 C_{x_1}}$, $\beta =\displaystyle\frac{1}{\chi
R_7 C_{x_2}}=\displaystyle\frac{1}{\chi R_6 C_{x_2}}$, $\beta
r=\displaystyle\frac{1}{\chi R_8 C_{x_2}}$, $\gamma =\displaystyle\frac{1}{\chi
R_9 C_{x_3}}$ and $\gamma E=\displaystyle\frac{1}{\chi R_{10} C_{x_3}}$
where $\chi$ is a parameter chosen as $\chi=10^4$. Thus, considering that
$R=10k\Omega$ and all capacitors are $C=10nF$, we
obtain the following components values $R_1=R_5=3988.095238\Omega$,
$R_3=188.2189546\Omega$, $R_2=3.48k\Omega$, $R_4=46261.31262\Omega$,
$R_6=R_7=33.5k\Omega$, $R_8=2093.75\Omega$, $R_9=47761.19403\Omega$ and
$R_{10}=191044.7761\Omega$.
The chaotic behavior of the above circuit is given by the
graphs of the following Fig.~(\ref{Fig.ppor}).
   \begin{figure}[htp]
 \begin{minipage}[b]{8cm}
      \begin{center}
    \includegraphics[scale=0.15]{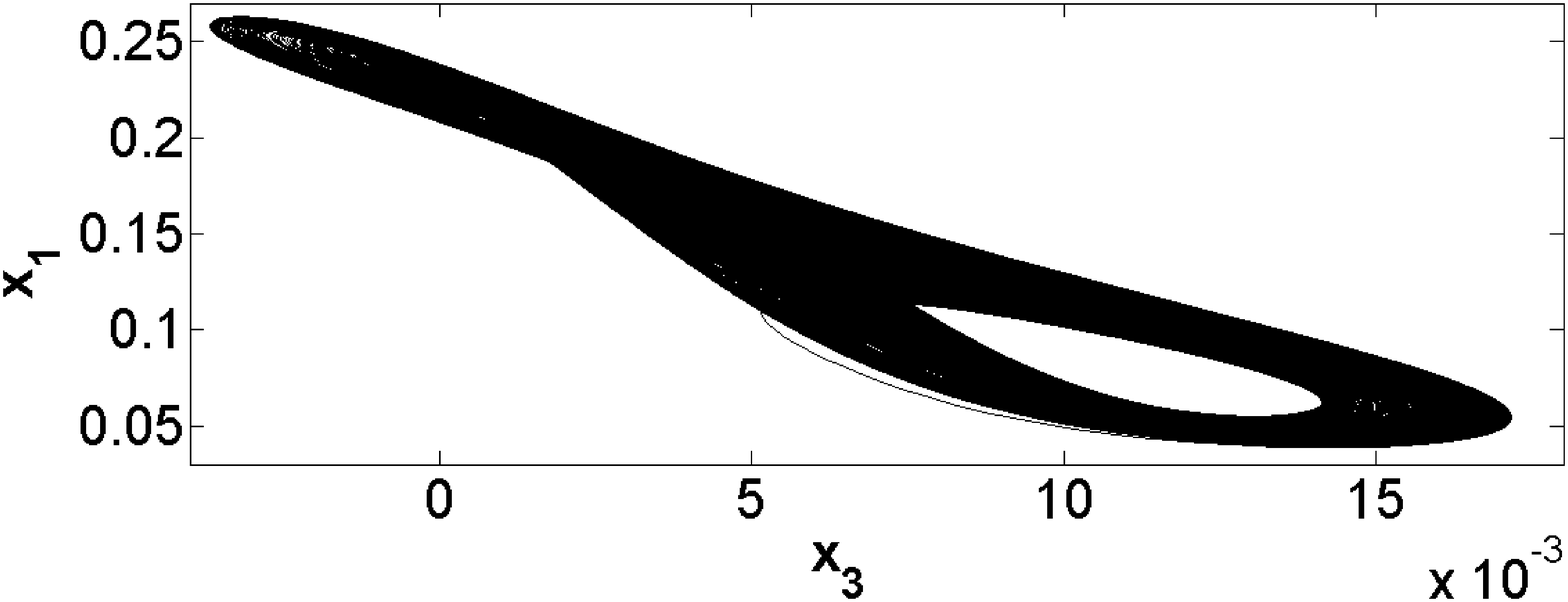}(a)
        \includegraphics[scale=0.15]{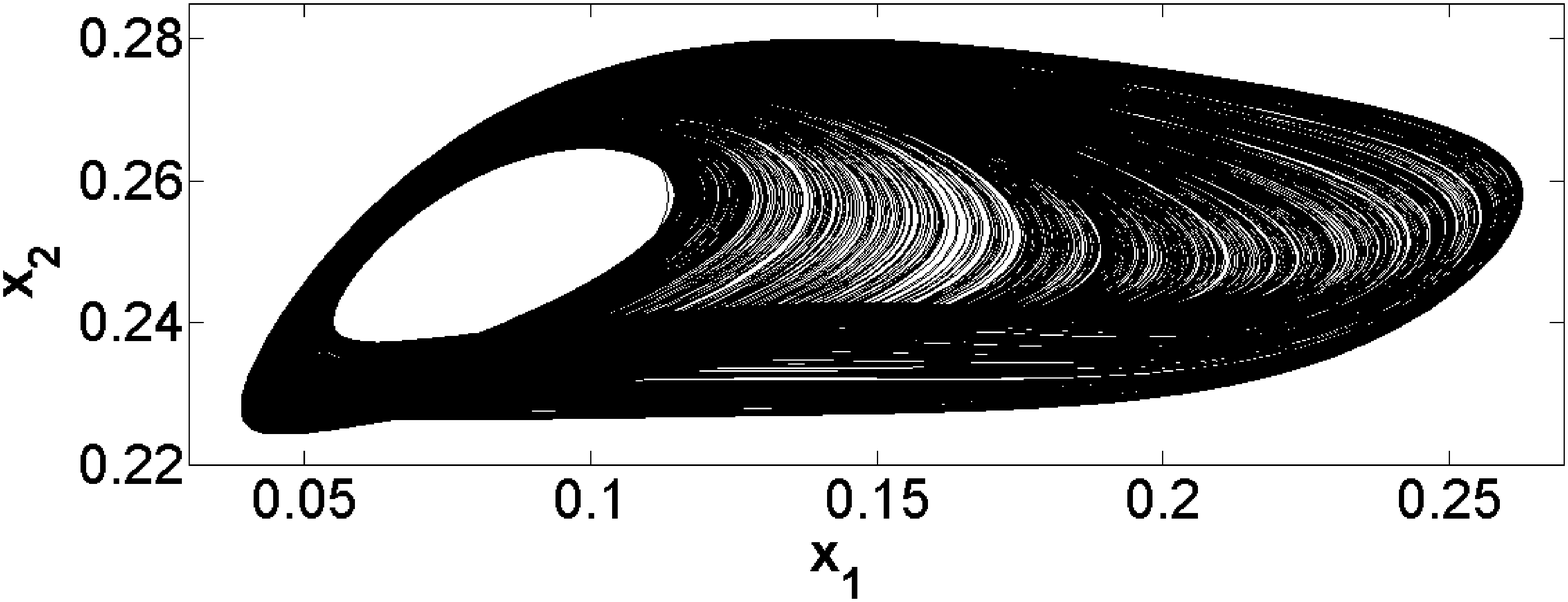}(b)
        \includegraphics[scale=0.15]{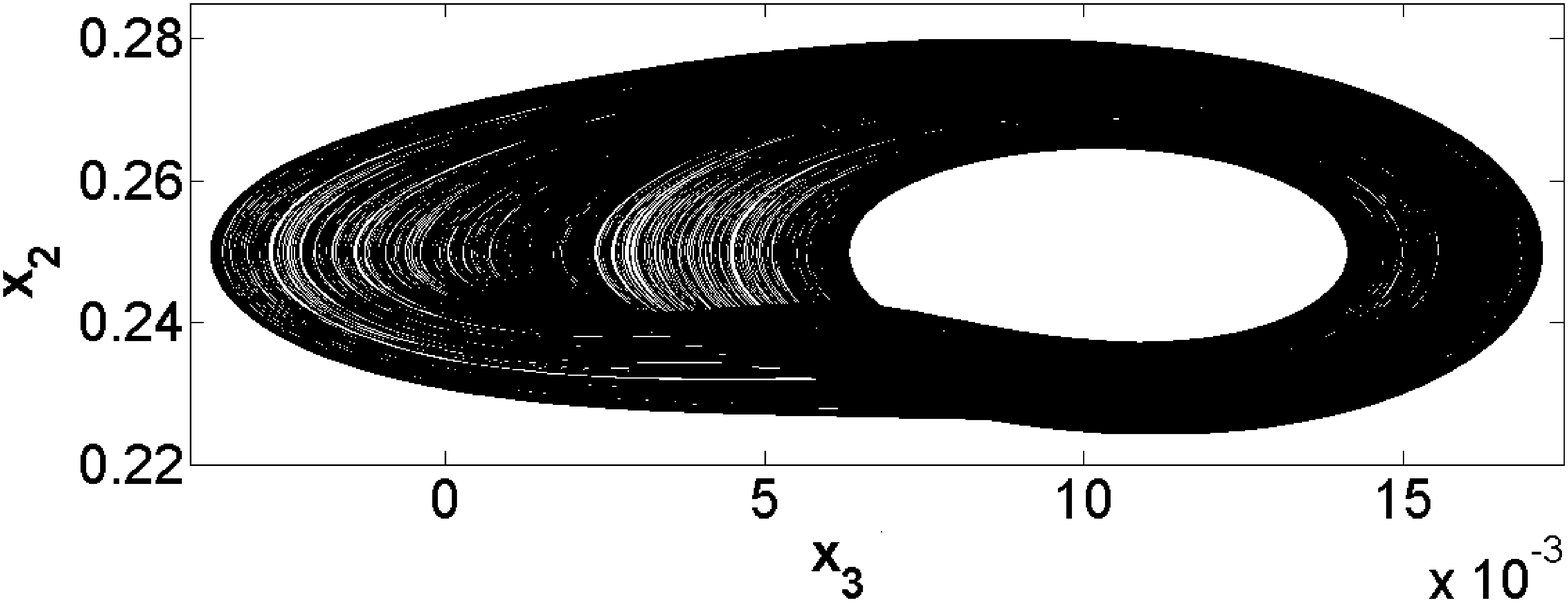}(b)
   \caption{Chaotic attractor from circuit in Fig.~(\ref{anlc})}
     \label{Fig.ppor}
     \end{center}
     \end{minipage}
     \end{figure}

\begin{figure}[htp]
 \begin{minipage}[b]{8cm}
      \begin{center}
        \includegraphics[scale=0.4]{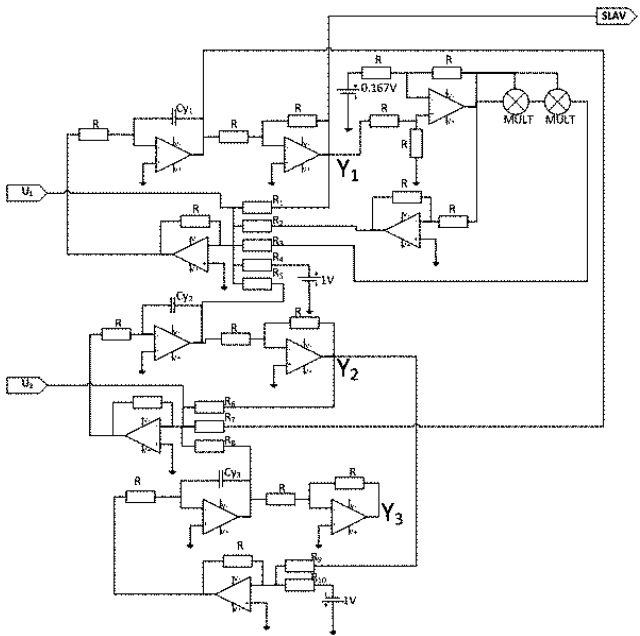}
   \caption{Circuit diagram of the response system.}
     \label{slave}
     \end{center}
     \end{minipage}
     \end{figure}

\begin{figure}[htp]
 \begin{minipage}[b]{8cm}
       \begin{center}
        \includegraphics[scale=0.4]{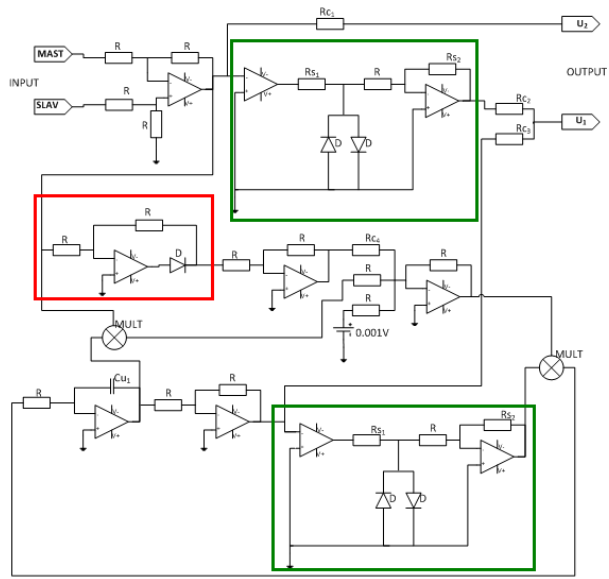}
   \caption{Circuit diagram of the controller}
     \label{pcont}
     \end{center}
     \end{minipage}
     \end{figure}

\begin{figure}[htp]
 \begin{minipage}[b]{8cm}
      \begin{center}
      \includegraphics[scale=0.15]{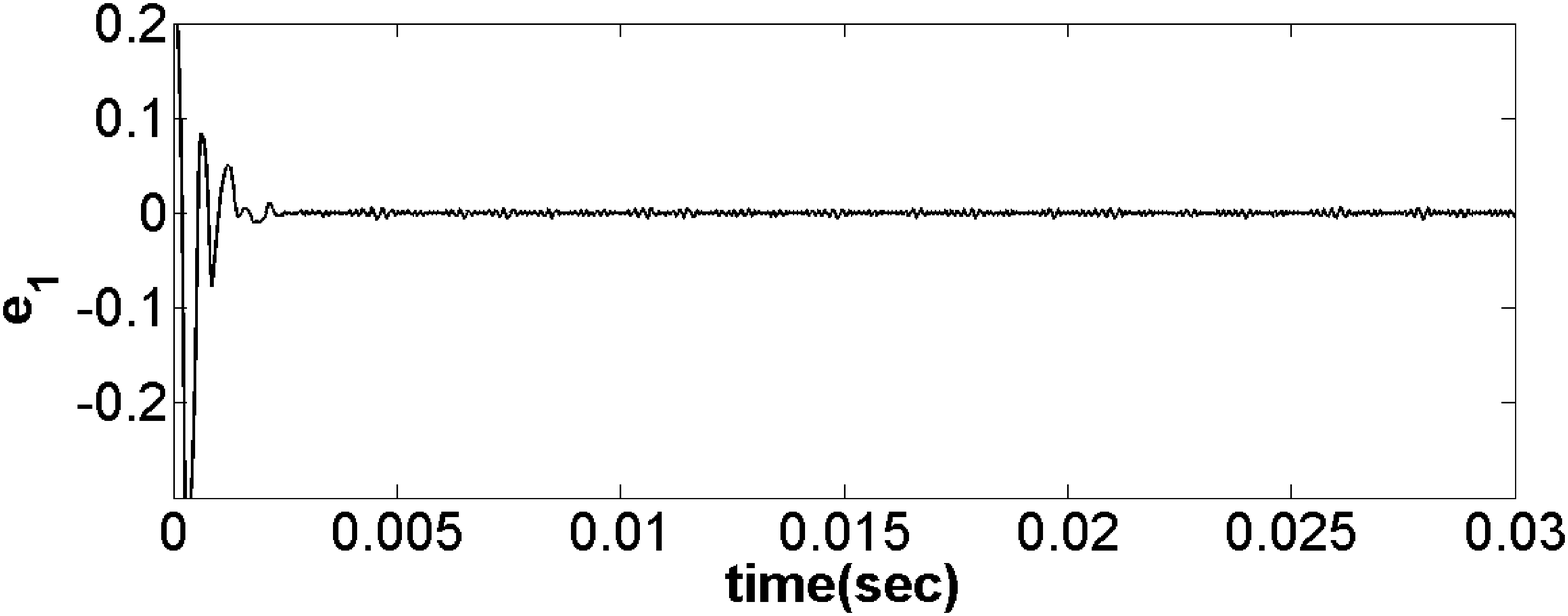}(a)
    \includegraphics[scale=0.15]{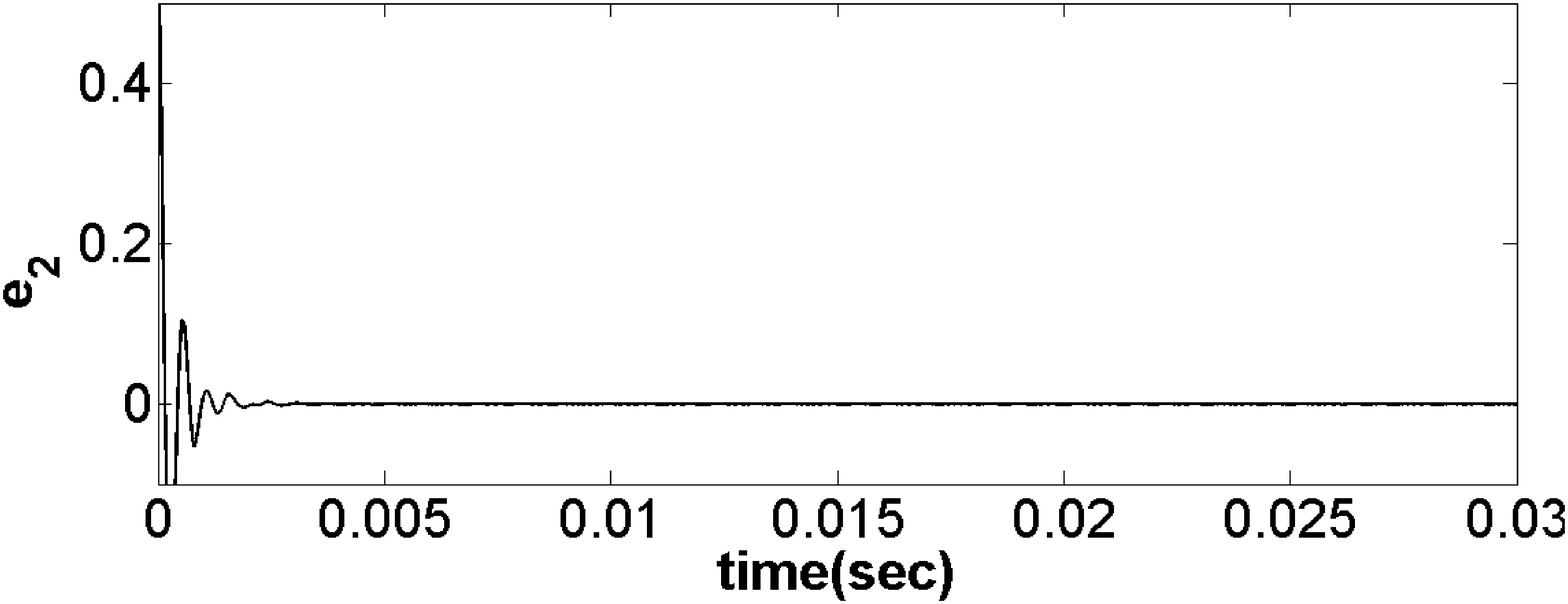}(b)
        \includegraphics[scale=0.15]{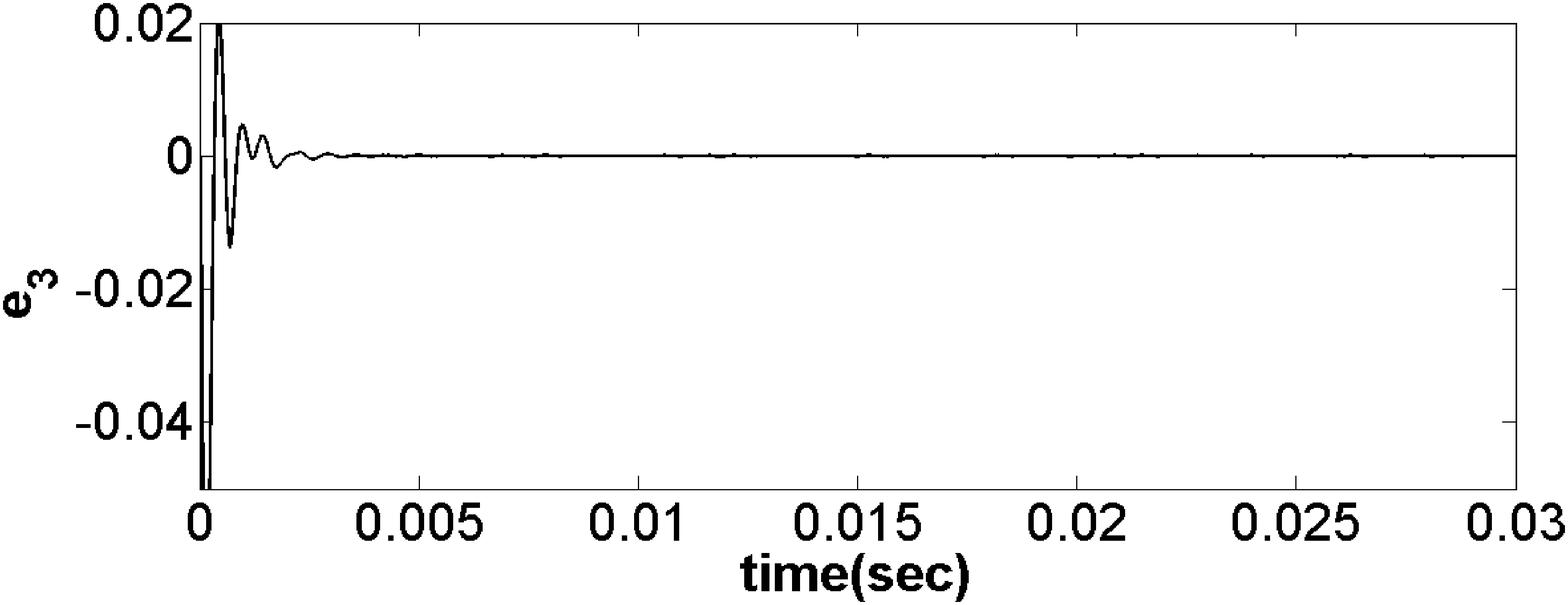}(c)
      \includegraphics[scale=0.2]{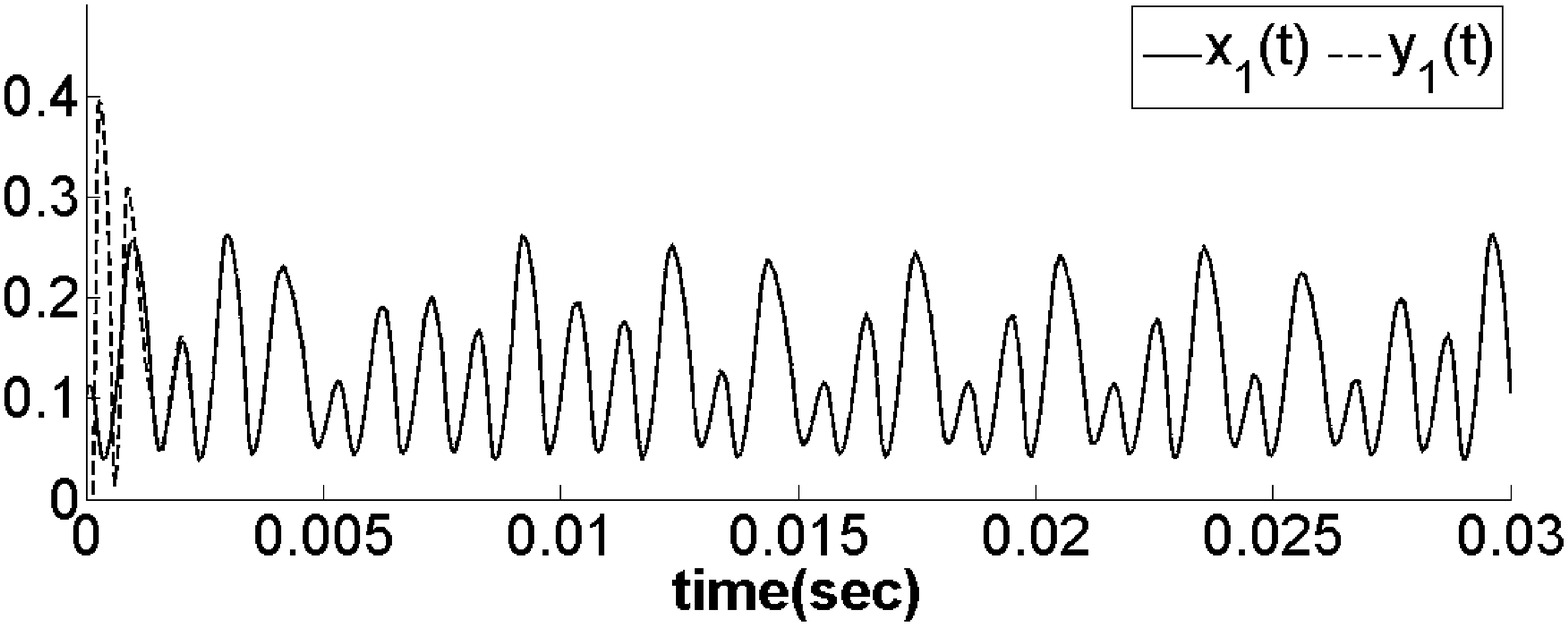}(d)
      \includegraphics[scale=0.2]{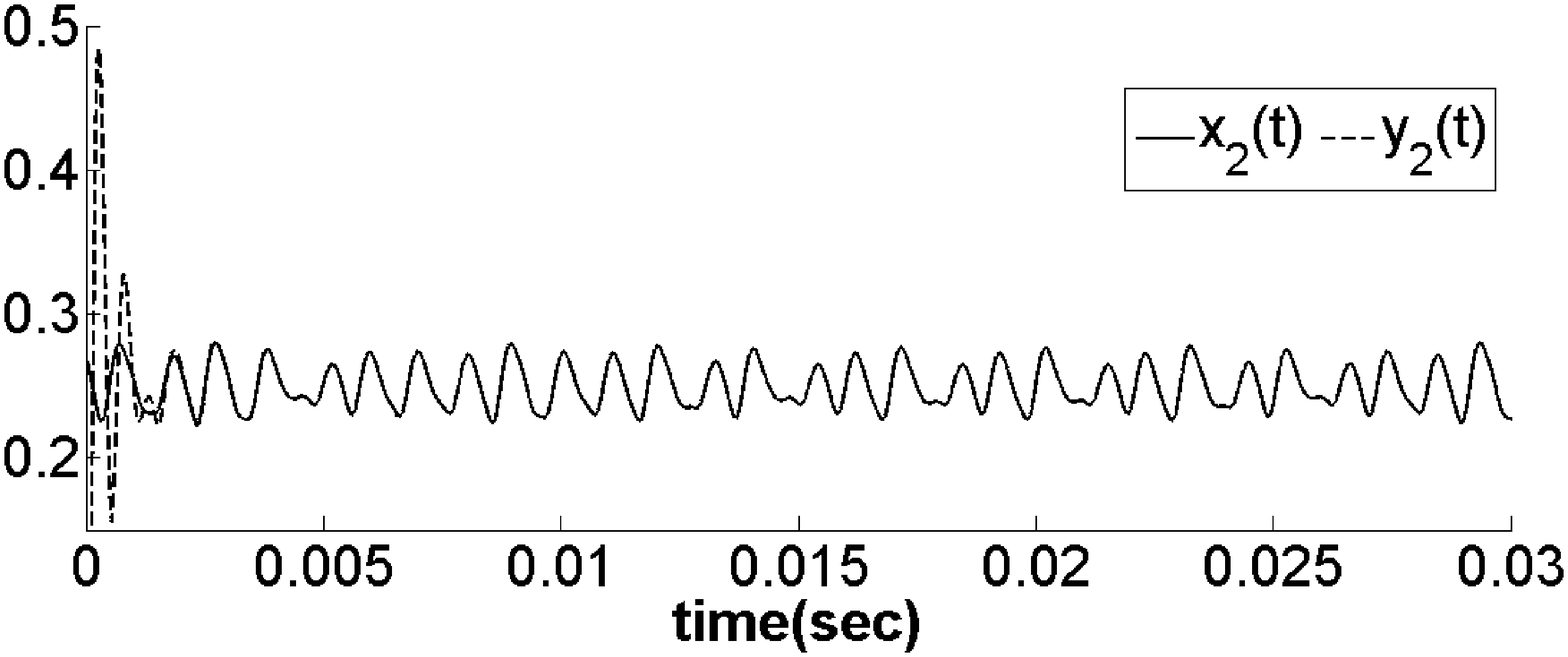}(e)
      \includegraphics[scale=0.2]{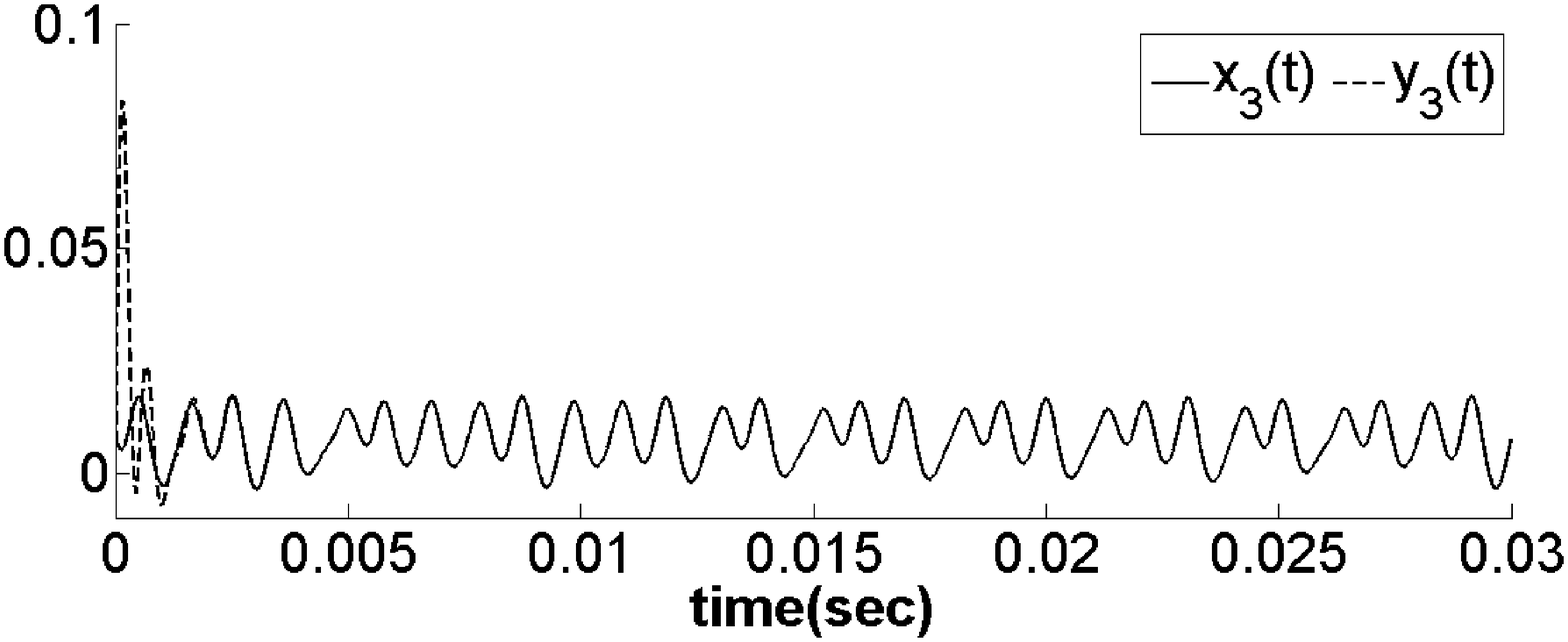}(f)

   \caption{Time histories of $e_i(t), i=1,2,3$ and of the systems variables.}
     \label{res1}
     \end{center}
     \end{minipage}
     \end{figure}

Later on the drive (Fig.~(\ref{drive})) and response (Fig.~(\ref{slave}))
systems are designed as
follows:
   \begin{figure}[htp]
 \begin{minipage}[b]{8cm}
      \begin{center}
        \includegraphics[scale=0.4]{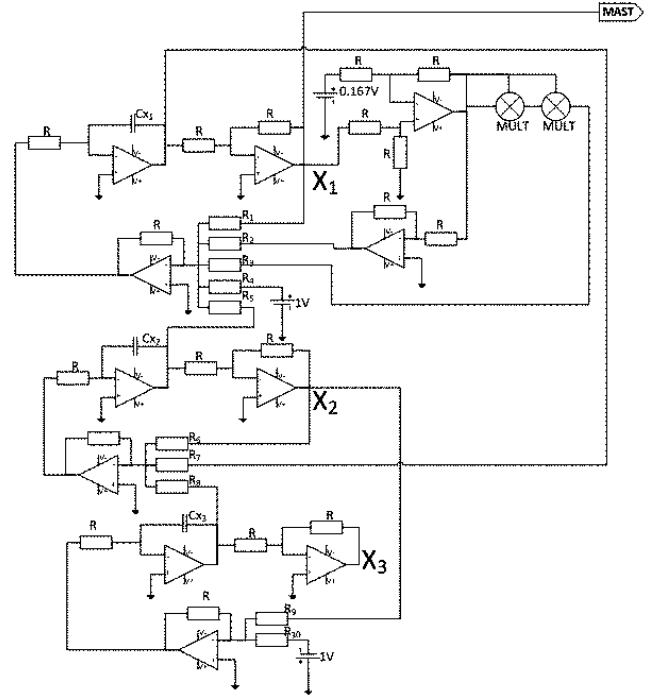}
   \caption{Circuit diagram of the drive system.}
     \label{drive}
     \end{center}
     \end{minipage}
     \end{figure}
%
%
%
The controller (See Fig.~(\ref{pcont})) is designed using the
following parameters: $2\beta=\displaystyle\frac{1}{\chi R_{C_1}C_{y_2}}$,
$\zeta
k =\displaystyle\frac{1}{\chi R_{C_2}
C_{y_1}}$, $\zeta =\displaystyle\frac{1}{\chi R_{C_3}
C_{y_1}}$, $k =\displaystyle\frac{1}{\chi R_{C_4} C_{u_1}}$, $1
=\displaystyle\frac{1}{\chi R C_{u_1}}$
and $p =\displaystyle\frac{0.001}{\chi R C_{u_1}}$ with which we obtain the
 components values , $R_{C_1}=16.75k\Omega$ ,
$R_{C_2}=797619.0476\Omega$, $R_{C_3}=3988.095238\Omega$, $R_{C_4}=2M\Omega$,
$R_{S_1}=1k\Omega$ and $R_{S_2}=13.508k\Omega$. The part of the controller
circuit in red box, according to J.C. Sprott~\cite{spr},
simulates the absolute value function while the one in green
domains simulates the sign function~\cite{lin}.
The  results obtained from these circuits are given in Fig.~(\ref{res1}).

\end{document}